\documentstyle[12pt,psfig]{article}
\input xy
\xyoption{all}

\newcommand\cO{\mathcal{O}}
\newcommand\cV{\mathcal{V}}

\newcommand\Hom{\mathrm{Hom}}

\newtheorem{defn}{Definition}
\newtheorem{lem}{Lemma}
\newtheorem{prop}{Proposition}

\begin{document}

\hfill ILL-(TH)-03-09

\hfill hep-th/0309270

\vspace{0.5in}

\begin{center}

{\large\bf Spectra of D-branes with Higgs vevs}

\vspace{0.2in}

Ron Donagi$^1$, Sheldon Katz$^{2,3}$, and Eric Sharpe$^2$ \\
$^1$ Department of Mathematics \\
David Rittenhouse Laboratory \\
University of Pennsylvania \\
209 South 33rd Street \\
Philadelphia, PA  19104 \\
$^2$ Department of Mathematics \\
1409 W. Green St., MC-382 \\
University of Illinois \\
Urbana, IL  61801 \\
$^2$ Department of Physics\\
University of Illinois at Urbana-Champaign\\
Urbana, IL 61801\\
{\tt donagi@math.upenn.edu}, {\tt katz@math.uiuc.edu},
{\tt ersharpe@uiuc.edu} \\

\end{center}

In this paper we continue previous work on counting open string states
between D-branes
by considering open strings between D-branes with nonzero Higgs vevs,
and in particular, {\it nilpotent} Higgs vevs,
as arise, for example, when studying D-branes in orbifolds.
Ordinarily Higgs vevs can be interpreted as moving the D-brane,
but nilpotent Higgs vevs have zero eigenvalues, and so their interpretation
is more interesting -- for example, they often correspond to nonreduced
schemes, which furnishes an important link in understanding old results
relating
classical D-brane moduli spaces in orbifolds to Hilbert schemes,
resolutions of quotient spaces, and the McKay correspondence.  
We give a sheaf-theoretic description of 
D-branes with Higgs vevs, including nilpotent Higgs vevs,
and check that description by noting
that Ext groups between the sheaves modelling the D-branes,
do in fact correctly count open string states.
In particular, our analysis expands the types of sheaves which 
admit on-shell physical interpretations,
which is an important step for making derived categories useful for physics.

\begin{flushleft}
October 2003
\end{flushleft}

\newpage

\tableofcontents

\newpage

\section{Introduction}

It was originally proposed in \cite{hm} that under certain
conditions, D-branes could be modelled mathematically by
sheaves.  Since that time, it has become popular to assume
that sheaves are good models for D-branes, and that many physical
properties of open strings can be calculated mathematically using
sheaves, assumptions that went unchecked until
relatively recently \cite{ks,kps,cks,merev}.  
It has even become popular to use
derived categories of sheaves to model off-shell states in string theory
(see \cite{paulron,medc,mikedc,paulalb} for an incomplete list
of early references).

However, one of the unresolved questions in all this work is
simply, do all sheaves correspond to D-branes?
Some special sheaves correspond to D-branes in a relatively
straightforward fashion -- for example, sheaves of the form
$i_* {\cal E}$ for inclusion map $i: S \hookrightarrow X$
correspond to D-branes wrapped on submanifold $S$ with
gauge bundle ${\cal E} \otimes \sqrt{K_S^{\vee}}$ \cite{ks,merev}.
But not all sheaves are of this form -- do the others have any
physical interpretation in terms of D-branes?

In light of \cite{paulron,medc,mikedc,paulalb}, the question
can be put differently.  For any sheaf, one can write down
a massive non-conformal, off-shell theory that formally corresponds to the
sheaf.  However, for which sheaves does the massive non-conformal theory
have a nontrivial RG fixed point?  For which sheaves is there an
on-shell description, and what is that on-shell description?

In this paper we answer this question in some new cases.
In particular, in this paper we describe how the data of
Higgs vevs can be encoded in sheaves.  For some (trivial) Higgs vevs,
the effect is to move the support of the sheaf.
For other (nontrivial) Higgs vevs, the effect is to create new sheaves,
that are not of the form of pushforwards of vector bundles,
but rather are things like, for example, structure sheaves
of nonreduced schemes.  Modelling D-branes with nilpotent
Higgs vevs with nonreduced schemes was first proposed in
\cite{tomasme}.  Here we are able to greatly clarify and extend
the ideas of that paper, and check the proposal by
directly computing open string spectra between
D-branes with nilpotent Higgs vevs, and showing that the open string
spectra are correctly counted by Ext groups between sheaves
on nonreduced schemes -- the proposed mathematical ansatz
correctly encodes physics.

Another issue is also addressed in this paper.
In \cite{dgm} it was shown that the classical Higgs moduli space
of D-branes on orbifolds involves a resolution of the quotient space,
a fact which has led some physicists to claim that string orbifolds
describe strings on resolutions on quotient spaces.
These resolutions were later shown to be precisely $G$-equivariant
Hilbert schemes of points on the covering spaces, in which
the exceptional divisors of the resolutions corresponded
(in the Hilbert scheme) to $G$-equivariant nonreduced schemes on
the covering space.  Now, although such correspondences were
checked algebraically, one could certainly ask {\it why} such
a correspondence should exist at all.

We are able to address this old puzzle in this paper.
The nonreduced schemes appearing in orbifolds correspond,
via the mathematical ansatz we shall describe herein,
to D-branes with nilpotent Higgs vevs, which are what
one physically computes on a D-brane in an orbifold background.

A third issue is also addressed here.
Whereas physically in \cite{dgm} one sees nilpotent Higgs vevs
mapping out exceptional divisors, the McKay correspondence 
in the form \cite{bkr} yields nonreduced schemes in place of
nilpotent Higgs vevs.  For consistency, one would like the two
to be closely related, and in this paper, we work out the 
precise relationship.

A fourth issue is also clarified in this paper.
The results of \cite{dgm} on classical moduli spaces of Higgs vevs
on D-branes in orbifolds have led some physicists to claim that
string orbifolds describe strings propagating on resolutions of
quotient spaces.  By pointing out that the exceptional divisors
are really nilpotent Higgs vevs, which are functionally equivalent
to $G$-equivariant nonreduced schemes, {\it i.e.} nonreduced
schemes on the quotient stack, we see how the results
of \cite{dgm} are completely consistent with the claim that
string orbifolds describe strings propagating on quotient stacks
\cite{qs}.  In fact, precisely this mechanism was used in
\cite{qs} to address this issue; the results in this paper
considerably clarify and strengthen this mechanism for resolving
the apparent contradiction.

We begin in section~\ref{brst} by describing the worldsheet
computation of open string spectra between D-branes with Higgs vevs.
The effect of the Higgs vevs is to modify the BRST operator
in a fashion that is by now relatively well-understood -- see
\cite{paulalb,calin} for other references using similar methods.
In section~\ref{mathansatz} we outline our mathematical ansatz
for encoding Higgs vevs in sheaves.  This ansatz previously appeared
in the math literature in \cite{del}, and its application to
D-branes was first proposed in \cite{tomasme}.  In appendix~\ref{pfs}
we prove that the massless spectra computations of section~\ref{brst}
always give the same result as Ext groups between the sheaves generated
by the ansatz in section~\ref{mathansatz}, but to help clarify
matters, we do describe a number of examples.
In section~\ref{check1} we describe a case with relatively trivial
Higgs vevs, which merely have the effect of moving the sheaf.
In section~\ref{d0exs} we describe more interesting examples
with nilpotent Higgs vevs, which cannot move the sheaf,
but rather generate sheaves which previously have not had an on-shell
physical interpretation.  In section~\ref{obs} we consider
sheaf models of 
Higgs vevs in D-branes wrapped on obstructed curves, which illuminate
some technical aspects of our work.

In section~\ref{fterms} we pause to outline how the physical F-term
conditions that one must impose on Higgs vevs in order to preserve
supersymmetry, are also necessary conditions for the corresponding
sheaf moduli to be unobstructed.

In section~\ref{orbs} we finally discuss one of the primary motivations
for this work, namely orbifolds.  Nilpotent Higgs vevs play a
very important role in orbifolds, as they are responsible for the exceptional
divisors of resolutions of quotient spaces seen in moduli 
spaces of classical Higgs vacua of D-branes on orbifolds.
We review this role of nilpotent Higgs vevs in orbifolds,
and also discuss how the McKay correspondence implies an alternative
interpretation of nilpotent Higgs vevs -- in terms of nonreduced schemes,
an interpretation supported by our results.
In section~\ref{orbexs} we study several examples of D-branes in
orbifolds with nonzero Higgs vevs, and compare to Ext groups between
the corresponding sheaves.

Throughout this paper we will make an important technical
assumption, namely we will only consider D-branes wrapped on
submanifolds $S$ with the property that
$TX|_S$ splits {\it holomorphically}
into $TS \oplus {\cal N}_{S/X}$.  This assumption is necessary
in order to guarantee that Higgs fields in the sense of
algebraic geometry, {\it i.e.} holomorphic sections of 
the normal bundle tensored with endomorphisms of the
gauge bundle, do indeed all correspond to zero modes of
D-brane fields.  As observed in \cite{ks,merev}, it is
{\it not} true that such Higgs fields always occur physically -- 
because motions of the brane are not always compatible with the
gauge bundle on the D-brane, sometimes some Higgs fields do
not represent infinitesimal moduli, and so do not appear physically.
By assuming that $TX|_S$ splits holomorphically as $TS \oplus {\cal N}_{S/X}$,
we are able to guarantee that Higgs fields in the algebraic
geometry sense above do appear physically, and so we shall make
this assumption throughout this paper.

Another consequence of this assumption is that the spectral
sequences that played an important role in \cite{ks,kps,cks,merev}
all trivialize.  Thus, there will be no spectral sequences
in this paper.

A short summary of the results of this paper appeared
previously in \cite{merev}.

\section{First-principles worldsheet analysis}   \label{brst}

What does turning on a Higgs vev do to open string spectrum calculations,
from the perspective of the worldsheet?

The first effect is to modify the BRST operator.
The most efficient way to see this is as follows.

Turning on a Higgs vev on the worldsheet $\Sigma$ in the open
string B model, following the standard procedure, means adding 
\begin{equation}  \label{boundinsert}
P \exp \int_{\partial \Sigma} \{ G, V \}
\end{equation}
to correlation functions,
where $G$ is the other half of the worldsheet boundary supersymmetry,
obeying $\{ Q, G \} \propto d$ with respect to the BRST charge $Q$,
and $V$ is the vertex operator for the Higgs field,
which is of the form
\begin{displaymath}
V \: = \: \Phi^{\alpha \beta i} \theta_i
\end{displaymath}
where $\theta_i$ are B model worldsheet fermions and
$\Phi^{\alpha \beta i}$ are the Higgs fields,
in the notation of \cite{ks}.
Deforming the action in this way has the effect of modifying the
BRST operator, as we shall see momentarily.

Before seeing how the BRST operator is deformed, let us first
pause for a moment to check that the prescription above gives
reasonable results.
If the vertex operator $V$ corresponded to a deformation of the
gauge field, {\it i.e.},
\begin{displaymath}
V \: = \: \left( \delta A_{\overline{\imath}} \right)
\eta^{ \overline{\imath}}
\end{displaymath}
then 
\begin{displaymath}
\{ G, V \} \: \propto \: \left( \delta A_{ \overline{\imath} } \right)
d \phi^{ \overline{\imath} } \: + \:
\partial_j \left( \delta A_{ \overline{\imath} } \right) \eta^{\overline{
\imath}} \rho^j
\end{displaymath}
which is clearly a deformation of the usual boundary gauge coupling
\begin{displaymath}
A_{\mu} d \phi^{\mu} \: + \: F_{j \overline{\imath} }
\eta^{\overline{\imath}} \rho^j
\end{displaymath}
In the present case, for the boundary state
\begin{displaymath}
V \: = \: \Phi^i(\phi) \theta_i
\end{displaymath}
we have
\begin{displaymath}
\{ G, V \} \: \propto \: g_{i \overline{\jmath} } \Phi^i 
(\partial - \overline{\partial} ) \phi^{ \overline{\jmath} }
\: + \: i ( D_j \Phi^i ) \rho^j \theta_i
\end{displaymath}
and, indeed, it is well-known that
the boundary coupling of Higgs fields is a 
supersymmetrization of the 
boundary interaction 
\begin{displaymath}
\int_{\partial \Sigma} d \tau \Phi_{\mu} (\partial - \overline{\partial}) \phi^{
\mu }
\end{displaymath}
for Higgs field $\Phi_{\mu}$.

Now, let us return to the question of the deformation of the BRST
operator. 
The quick way to see how the BRST operator is deformed is as follows
(see also \cite{paulalb,calin} for other instances of this derivation).
Recall that correlation functions in the deformed BCFT are defined
by inserting~(\ref{boundinsert}) in correlation functions in the original
BCFT, so, for example,
\begin{displaymath}
< \{ Q, V \} >_{new} \: = \:
< \left( P \exp \int_{\partial \Sigma} \{ G, V_1 \} \right)
\{ Q, V \}
 \left( P \exp \int_{\partial \Sigma} \{ G, V_2 \} \right)
>_{old}
\end{displaymath}
Now, let us work out what it means for $V$ to be annihilated by
the BRST operator in the deformed theory.

Commuting the BRST operator $Q$ past the insertions brings down
factors proportional to $\int_{\partial \Sigma} \{ Q, G \} V_i$,
and using the fact that $\{ Q, G \} \propto d$, we see that this has
the effect of adding terms $V_i$ from the boundaries of the
integrals $\int_{\partial \Sigma} d V_i$:
\begin{displaymath}
< \{ Q_{new}, V\} >_{new} \: = \: 0
\mbox{ if and only if }
\{ Q_{old} \: + \: V_1 \: - \: V_2, V \} \: = \: 0
\end{displaymath}
As a result, we identify
\begin{displaymath}
Q_{new} \: = \: Q_{old} \: + \: V_2 \: - \: V_1
\end{displaymath}
where $V_1$, $V_2$ are the vertex operators of the form
\begin{displaymath}
V_m \: = \: \Phi^i_m \theta_i
\end{displaymath}
describing Higgs vevs on either end of an open string.

In order to preserve ${\cal N}=2$ supersymmetry, we must demand
that $Q^2=0$.
It is easy to show that
\begin{displaymath}
Q^2 \cdot V \: = \: \left( \overline{\partial} \Phi_1^i \right) V \theta_1
\: - \: V \left( \overline{\partial} \Phi_2^i \right) \theta_i
\: + \: \left( \Phi_1^j \Phi_1^i V \right) \theta_j \theta_i \: + \:
\left( V \Phi_2^i \Phi_2^j \right) \theta_j \theta_i
\end{displaymath}
where we have identified $Q_{old} = \overline{\partial}$.
Clearly, in order for $Q^2$ to vanish, we must demand that the
$\Phi$'s define holomorphic sections of the normal bundle (tensored with
endomorphisms of the gauge bundle), and that the $\Phi$'s associated with
any one D-brane commute with one another.

Commutativity of Higgs vevs is a typical F-term constraint on
supersymmetric vacua, as we shall see explicitly in discussions
of orbifolds in section~\ref{dbraneorbcomp}.
In addition, in a {\it physical} theory there are D-term constraints,
which ordinarily exclude nilpotent Higgs vevs, with the important
exception of orbifolds, where nilpotent Higgs vevs play an important role.
However, we are considering the topological B model, not a physical theory,
and so there are no D-term conditions to worry about here.

Thus, given a set of commuting Higgs vevs $\Phi^i$
on each of a pair of sets of
D-branes, and for the moment ignoring twistings of boundary conditions
induced by Chan-Paton curvature, we see that the open string spectrum
is given by cohomology of the deformed BRST operator
\begin{displaymath}
Q \: = \: \overline{\partial} \: + \: \Phi_i^i \theta_i \: - \:
\Phi_2^i \theta_i
\end{displaymath}

We shall explicitly compute open string spectra using these
methods in a number of examples throughout this paper.
Now that we have outlined how one calculates the physical massless
boundary Ramond sector spectrum between D-branes with Higgs vevs,
we shall next describe the mathematical ansatz for encoding
such Higgs vevs in sheaves.  We shall check that that mathematical
ansatz is physically relevant by showing that the physical massless
spectrum, computed as in this section, is correctly counted by
Ext groups between sheaves generated by the mathematical ansatz.

\section{Sheaves from Higgs vevs}   \label{mathansatz}

Ultimately we want to relate nonzero Higgs vevs to sheaves,
so as, for example, to be able to simplify computations of open string
spectra by turning them into mathematical computations of Ext groups.
In fact, there is a relatively straightforward method to do exactly this.

For simplicity, let us for the moment assume that the Chan-Paton
factors have no curvature, so as to simplify the boundary conditions
on worldsheet fields.  In this case, the Higgs field is merely
an adjoint-valued section of the normal bundle to the submanifold.
The key to finding a sheaf-y description of a Higgs vev is to
interpret the Higgs field as deforming the action of the coordinate
ring on the module describing the sheaf, just as in \cite{del}.
(See also \cite{tonysb} for some recent lectures on this
bit of mathematics, and for related comments on Higgs fields
in the A model.)

A detailed explanation is necessarily algebraic.
We beg the reader's indulgence for a moment while we briefly outline 
this construction.  As described above, a Higgs field $\Phi$
in the sense of algebraic geometry
is an adjoint-valued holomorphic section of the normal bundle:
\begin{displaymath}
\Phi \: = \: H^0\left(S, {\cal N}_{S/X} \otimes {\cal E}^{\vee} \otimes {\cal E}
\right)
\end{displaymath}
In order to understand more precisely how the Higgs field can
deform the ring action, we shall rewrite this description:
\begin{displaymath}
 H^0\left(S, {\cal N}_{S/X} \otimes {\cal E}^{\vee} \otimes {\cal E}\right)
\: = \: \mbox{Hom}\left( {\cal N}_{S/X}^{\vee}, {\cal E}^{\vee} \otimes {\cal E} \right)
\end{displaymath}
and the conormal bundle ${\cal N}_{S/X}^{\vee} = {\cal I}/{\cal I}^2$,
where ${\cal I}$ is the ideal defining the subvariety $S$,
{\it i.e.}, locally, $X = \mbox{Spec }A$ and $S = \mbox{Spec } A/{\cal I}$.
Now, a sheaf on $X$ supported on $S$ is an $A$-module $M$
which is annihilated by any element of ${\cal I}$.
We can create a new $A$-module, and hence a new sheaf on $X$,
by starting with the original module $M$ and deforming the $A$ action
so that, schematically, for any $x \in {\cal I}$,
\begin{displaymath}
x \cdot M \: = \: \Phi(x) \cdot M
\end{displaymath}
where $\Phi$ is the Higgs field,
here interpreted as a homomorphism from ${\cal I}/{\cal I}^2$
to automorphisms of the vector bundle on $S$ described by the
module $M$.
If the section is identically zero -- if the Higgs fields all vanish
 -- then the new module is identical to the original one.

This mathematical construction is described in more detail
in appendix~\ref{pfs}, but as this paper is intended for a physics
audience, for the moment we content ourselves with the outline above.

Now, how can we tell if the mathematical construction outlined
above has any physical relevance?  This construction has physical
relevance precisely if physical properties of D-branes with
Higgs vevs can be calculated mathematically using the
sheaves constructed above.  Indeed, we prove in appendix~\ref{pfs}
that massless boundary Ramond sector spectra between D-branes
with Higgs vevs, as computed in the previous section,
are also always computed by Ext groups between the sheaves
obtained by the mathematical construction just outlined.
Thus, our mathematical ansatz does indeed have physical content,
as we shall see explicitly in numerous examples computed elsewhere
in this paper.

With this description in mind, we can also quickly appreciate
the implicit mathematical content of
the BRST operator described in the previous section.
Consider a degree zero state $V$.
Then, the condition that $Q \cdot V = 0$ factorizes into several components:
\begin{eqnarray*}
\overline{\partial} V & = & 0 \\
\Phi^i_1 V & = & V \Phi^i_2 \mbox{ for each }i
\end{eqnarray*}
where $\Phi^i_1$, $\Phi^i_2$ are the components of the Higgs
field for the two D-branes.  With our description of modified ring
actions above in mind, the second condition is just the statement
that $V \in \mbox{Ext}^0$ defines a module homomorphism, 
as elements of $\mbox{Ext}^0 = \mbox{Hom}$ should -- the second condition
is just the statement of compatibility with the ring action.
Interpretation of the conditions on higher-degree states is more
subtle; precise equivalence of BRST cohomology with Ext group
elements is explained in appendix~\ref{pfs}.

Intuitively, what sheaves does the mathematical construction
above generate?
If the Higgs fields are, for example, diagonalizable with distinct
eigenvalues, then the effect of the deformation above is to
create a new sheaf with support over positions defined by the eigenvalues,
as one would expect physically.

On the other hand, if the Higgs fields are nilpotent, then something
much more interesting happens.  Since the eigenvalues of such fields
vanish, the D-branes do not move, but rather are modelled by,
for example, structure sheaves of non-reduced schemes.
The point that non-reduced schemes should describe D-branes with
nilpotent Higgs vevs has been made previously in the literature
by one of the authors \cite{tomasme} based on a close reading of
\cite{del}; here we shall be able to check that description in
detail, by explicitly comparing open string spectra in backgrounds
with nilpotent Higgs vevs to Ext groups between the corresponding sheaves.

Nilpotent Higgs fields are not often discussed in the physics literature.
One reason is that in typical cases, they are excluded by D-terms.
However, there are certainly physical situations in which they can appear.
For example, in studies of D-branes in orbifolds \cite{dgm},
the exceptional divisors resolving classical moduli spaces of Higgs
vevs are composed of nilpotent Higgs vevs in the original theory.
Later in section~\ref{dbraneorbcomp} we shall review the construction
of \cite{dgm} and review how nilpotent Higgs vevs arise in that context.
In any event, in the present case, there is a better reason to ignore D-terms,
as was mentioned in the previous section -- there is no analogue of
the D-term constraint in the topological theories we are considering.

So far in this section
we have described
a mathematical ansatz for encoding Higgs vevs in sheaves.
We prove in appendix~\ref{pfs} that this 
mathematical ansatz has physical content
by showing that the physical massless state computation
(as described in the previous section)
always gives the same result as Ext groups between 
the sheaves generated by the mathematical ansatz discussed
here.  

To clarify and explain our results, next we shall work through
a number of examples.  We begin with an easy example showing
how non-nilpotent Higgs vevs have the effect of moving the support
of the sheaf, as one would naively expect,
then later we consider some examples of nilpotent Higgs vevs and check
their sheaf-y description by comparing massless spectra to
Ext groups between the corresponding sheaves.

\section{Easy consistency check:  separable D0 branes}  \label{check1}

Let us perform an easy consistency check of our methods.
Non-nilpotent Higgs fields are interpreted as moving the D-branes
by a finite distance, and we know that if two D-branes do not intersect,
then there are no massless open string states between them.

So, as an easy test of our methods, let us start with an open
string stretching from one D0 brane at the origin of ${\bf C}$ to
another D0 brane, also at the origin of ${\bf C}$, and 
turn on a Higgs vev on one of the D0 branes.  That should be equivalent
to moving that D0 brane away from the origin, {\it i.e.} away from
the other D0 brane, and should remove any massless states from
the spectrum.

We shall begin by computing the physical massless spectrum,
and then we shall compare to Ext groups between the sheaves generated
by the mathematical ansatz.

In our conventions, the BRST operator $Q = \overline{\partial}
+ \Phi^i_1 \theta_i - \Phi^i_2\theta_i$ now simplifies
to become merely $\Phi \theta$, where $\Phi$ is the value of the nonzero Higgs
vev, and the single $\theta$ corresponds to the single
complex normal direction.

First, consider degree zero states.
These have the form
\begin{displaymath}
V \: = \: \alpha
\end{displaymath}
for some constant $\alpha$.
The BRST operator acts as, multiplication by $\Phi$,
so to be in the kernel of $Q$ means $\Phi \alpha = 0$.
Since $\Phi$ is nonzero by assumption, we have that
$\alpha = 0$, hence there are no degree zero states,
unless $\Phi=0$.

Next, consider degree one states.
These have the form
\begin{displaymath}
V \: = \: \alpha \theta
\end{displaymath}
for some constant $\alpha$.
The BRST operator annihilates all of these states, so they are
all in the kernel; but they are all in the image of the BRST operator
also, unless $\Phi=0$.  So, when $\Phi \neq 0$, we see that there
are no degree one states either.

Thus, in this example 
we have confirmed within our formalism the easy consistency
check that if we move D0 branes apart a nonzero distance by
turning on a Higgs vev, then there are no massless open string
states between them.

Next, let us compare to the sheaf-theoretic version
of Higgs fields we have just discussed.
{}From our general analysis outlined previously,
we are starting with a skyscraper sheaf at the origin of ${\bf C}$,
which is to say, a ${\bf C}[x]$-module with a single generator
$\alpha$, which is annihilated by $x \in {\bf C}[x]$,
and we are deforming the ring action on the module by defining
$x \cdot \alpha = \Phi \alpha$.
In other words, the new module also has a single generator,
which is annihilated by $(x-\Phi)$ -- which is to say,
the new module is the same as the skyscraper sheaf over
the point $x = \Phi$.

Thus, we see that our general analysis of how Higgs fields modify
sheaves predicts that turning on the Higgs field in this case
moves the skyscraper sheaf away from the origin, in exactly the form
one expects.

In general, however, this intuition can be slightly misleading.
Later we will consider an example of a rigid ${\bf P}^1$ in a 
Calabi-Yau three-fold, and despite the fact that ${\bf P}^1$ cannot move,
we will still be able to turn on a constant Higgs field.
The subtlety we are glossing over is that our construction of
sheaves from Higgs fields gives us sheaves in the normal bundle,
which need not be the same holomorphically as a local patch on the
Calabi-Yau.  We shall return to this issue in section~\ref{obs}.

\section{Examples of coincident D0 branes on points}  \label{d0exs}

In the last section we gave an example illustrating how
for `ordinary' Higgs vevs, our mathematical ansatz described
has the effect of moving the sheaf, as one would expect,
and checked that the physical massless spectrum computation
agrees with Ext groups between the sheaves generated by our ansatz.

In this section, we shall consider examples with much more
interesting Higgs vevs, namely nilpotent Higgs vevs.
Since their eigenvalues all vanish, they cannot be interpreted
as moving the sheaf, but rather `twist' the sheaf into something
not of the form of a pushforward of a vector bundle.
This will implicitly give us a new class of sheaves for which
physical meanings can be assigned, and we shall explicitly
check that physical meaning in each case by verifying that Ext groups count
massless states (specializing the general proof in appendix~\ref{pfs}).

In particular, our examples in this section will, for simplicity,
all involve
D0 branes on ${\bf C}^2$, sitting at the origin, with variable
Higgs vevs.  In each case, our BRST cohomology computations will reduce
to easy matrix manipulations.

Let us define some notation that will simplify our presentation.
Following \cite{del}, we will denote the skyscraper sheaf supported
at the origin by $C$ in the following subsections, rather than
${\cal O}_p$, and the direct sum of $n$ copies of the skyscraper
sheaf, {\it i.e.} ${\cal O}_p^n$, we shall for simplicity denote
by $nC$.  We shall use $D_x$ to denote the sheaf on ${\bf C}^2$
associated to the module ${\bf C}[x,y]/(x^2,y)$, {\it i.e.}
the structure sheaf of a (length two) nonreduced scheme supported at the origin.
We shall see in section~\ref{2cdx} that the sheaf $D_x$ corresponds
under our mathematical ansatz to a pair of D0 branes on ${\bf C}^2$
with Higgs fields
\begin{displaymath}
\Phi^x \: = \: \left[ \begin{array}{cc}
                      0 & 1 \\
                      0 & 0  \end{array} \right], \: \:
\Phi^y \: = \: \left[ \begin{array}{cc}
                      0 & 0 \\
                      0 & 0  \end{array} \right]
\end{displaymath}
Finally, we shall use $F$ to denote the sheaf associated to the
module ${\bf C}[x,y]/(x^2,xy,y^2)$, {\it i.e.} the structure sheaf
of a nonreduced scheme of length three supported at the origin.
We shall see in section~\ref{fc} that the sheaf $F$ corresponds
under our mathematical ansatz to three D0 branes on ${\bf C}^2$
with Higgs fields
\begin{displaymath}
\Phi^x \: = \: \left[ \begin{array}{ccc}
                      0 & 1 & 0 \\
                      0 & 0 & 0 \\
                      0 & 0 & 0  \end{array} \right], \: \:
\Phi^y \: = \: \left[ \begin{array}{ccc}
                      0 & 0 & 1 \\
                      0 & 0 & 0 \\
                      0 & 0 & 0  \end{array} \right]
\end{displaymath}

\subsection{No Higgs vevs on either side: $(2C, 2C)$}  \label{2c2c}

For our first, and easiest, example, we shall consider open strings
stretched between two pairs of D0 branes, all sitting at the origin
of ${\bf C}^2$, in the special
case that there are no Higgs vevs on either side of the open string.
This computation is a special case of that described in
\cite{ks}, but we review it for completeness.

Let us review the massless boundary Ramond sector state computation.
For open strings mapping a D-brane on a point back into itself,
the massless boundary Ramond sector states are of the form 
\cite{ks,merev}
\begin{displaymath}
b^{\alpha \beta j_1 \cdots j_m} \theta_{j_1} \cdots
\theta_{j_m}
\end{displaymath}
where the $\theta$ are B model worldsheet fields, and $\alpha$,
$\beta$ are Chan-Paton factors.
For open strings connecting a pair of D0-branes to a pair of D0-branes,
each $b^{\alpha \beta}$ is a $2 \times 2$ matrix.
Thus, we can write degree zero states in the form
\begin{displaymath}
V \: = \: \left[ \begin{array}{cc}
                 a & b \\
                 c & d \end{array} \right],
\end{displaymath}
degree one states in the form
\begin{displaymath}
V \: = \: \left[ \begin{array}{cc}
                 a_1 & b_1 \\
                 c_1 & d_1 \end{array} \right] \theta_1 \: + \:
\left[ \begin{array}{cc}
       a_2 & b_2 \\
       c_2 & d_2  \end{array} \right] \theta_2,
\end{displaymath}
and degree two states in the form
\begin{displaymath}
V \: = \: \left[ \begin{array}{cc}
                 a & b \\
                 c & d  \end{array} \right]
\end{displaymath}
Since these D-branes are in ${\bf C}^2$, there can be
no higher-degree states.  
Since these D-branes are wrapped on a point,
the BRST operator $Q = \overline{\partial}$ is completely trivial,
and so we see there are four states in degree zero, eight states in
degree one, and four states in degree two.

{}From the results above, we require that
\begin{displaymath}
\mbox{dim }\mbox{Ext}^n_{ {\bf C}^2 } \left( 2C, 2C \right) \: = \:
\left\{ \begin{array}{cl}
        4 & n=0 \\
        8 & n=1 \\
        4 & n=2
        \end{array} \right.
\end{displaymath}
which is easy to check is a correct mathematical statement.

\subsection{Higgs vevs on only one side:  $(2C, D_x)$}  \label{2cdx}

Here we shall consider open strings between two pairs of D0 branes,
all at the origin of ${\bf C}^2$,
with nonzero Higgs vevs on only one side of the open string.
We shall take the Higgs fields to be
\begin{displaymath}
\Phi^x \: = \: \left[ \begin{array}{cc}
                       0 & 1 \\
                       0 & 0
                       \end{array} \right],
\: \:
\Phi^y \: = \: \left[ \begin{array}{cc}
                      0 & 0 \\
                      0 & 0
                      \end{array} \right]
\end{displaymath}
If we start with the (trivial) rank 2 vector bundle on a point
in ${\bf C}^2$,
and deform the action of the ambient ring by the Higgs fields above,
then our bundle becomes the structure sheaf of the nonreduced scheme
of order 2 at the origin, defined by the ideal $(x^2,y)$,
and denoted $D_x$.  Thus, our open string computation should
be correctly reproduced by $\mbox{Ext}^*_{ {\bf C}^2} ( 2C, D_x )$.

First, consider degree zero states.
These are just matrices
\begin{displaymath}
V \: = \: \left[ \begin{array}{cc}
       a & b \\
       c & d 
       \end{array} \right]
\end{displaymath}
Demanding that the state above be in the kernel of $Q = \overline{\partial}
+ \Phi^i_1 \theta_i - \Phi^i_2 \theta_i $ means that
$V \Phi^x = 0$, so the $V$ in the kernel of $Q$ have the form
\begin{displaymath}
\left[ \begin{array}{cc}
       0 & b \\
       0 & d 
       \end{array} \right]
\end{displaymath}
Since there is no image to mod out, we see that the space of degree zero
states has dimension two.

Next, consider degree one states.
These can be written in the form
\begin{displaymath}
V \: = \: \left[ \begin{array}{cc}
                 a_x & b_x \\
                 c_x & d_x 
                 \end{array} \right] \theta_1 \: + \:
\left[ \begin{array}{cc}
       a_y & b_y \\
       c_y & d_y 
       \end{array} \right] \theta_2 
\end{displaymath}
States in the kernel of the BRST operator $Q$ can be written
\begin{displaymath}
V \: = \: \left[ \begin{array}{cc}
                 a_x & b_x \\
                 c_x & d_x 
                 \end{array} \right] \theta_1 \: + \:
\left[ \begin{array}{cc}
       0 & b_y \\
       0 & d_y 
       \end{array} \right] \theta_2 
\end{displaymath}
and states in the image of $Q$ can be written in the form
\begin{displaymath}
\left[ \begin{array}{cc}
       0 & a \\
       0 & c 
       \end{array} \right] \theta_1
\end{displaymath}
for some $a$, $c$.  Since the kernel is six-dimensional,
and the image is two-dimensional, we see that the space of
BRST-closed degree one states, modulo BRST exact states,
has dimension four.

Next, consider degree two states.
These can be written in the form
\begin{displaymath}
V \: = \: \left[ \begin{array}{cc}
                 a & b \\
                 c & d 
                 \end{array} \right] \theta_1 \theta_2
\end{displaymath}
The BRST operator $Q$ annihilates all of the degree two states,
and it is straightforward to check that the image of $Q$ in
degree two states has the form
\begin{displaymath}
\left[ \begin{array}{cc}
       0 & \alpha \\
       0 & \beta 
       \end{array} \right] \theta_1 \theta_2
\end{displaymath}
for some $\alpha$, $\beta$.
Hence the kernel of $Q$ has dimension four, and the image of $Q$
has dimension two, so the space of BRST-closed states, modulo
BRST-exact states, has dimension two.

Now, in order to interpret the results of these calculations
as coming from Ext groups between sheaves, we must find a sheaf-theoretic
interpretation of the pair of D0-branes with nontrivial Higgs fields.
Since there are two D0 branes, we have a module with two generators,
say, $\alpha$ and $\beta$.  The Higgs field associated to $x$
maps 
\begin{displaymath}
\left[ \begin{array}{c}
       \alpha \\ \beta \end{array} \right] \: \mapsto \:
\left[ \begin{array}{c}
       \beta \\ 0 \end{array} \right]
\end{displaymath}
so $x \cdot \alpha = \beta$ and $x \cdot \beta = 0$.
The Higgs field associated to $y$ annihilates both generators:
$y \cdot \alpha = x \cdot \beta = 0$.
Such a module is precisely ${\bf C}[x,y]/(x^2,y)$,
where we identify $\alpha$ with the image of $1 \in {\bf C}[x,y]$
in the quotient, and $\beta$ with the image of $x$ in the quotient.

This module defines an example of a nonreduced scheme.
It is supported at the origin, and has length two, but is not the
same as two copies of the skyscraper sheaf.
In fact, there is a ${\bf P}^1$'s worth of ideals of length two
by which we could quotient -- $D_x$ is not the only example of
a nonreduced scheme of length two supported at the origin of ${\bf C}^2$.

Thus, our computations predict
\begin{displaymath}
\mbox{dim }\mbox{Ext}^n_{ {\bf C}^2 } \left( 2C, D_x \right) \: = \:
\left\{ \begin{array}{cl}
         2 & n=0 \\
         4 & n=1 \\
         2 & n=2
         \end{array} \right.
\end{displaymath}

In fact, this result is easy to check.
First, Serre dualize to $\mbox{Ext}^*\left( D_x, 2C\right)$,
so that we can use the projective resolution of $D_x$,
given by
\begin{displaymath}
0 \: \longrightarrow \: {\cal O}_{ {\bf C}^2 }
\: \stackrel{{\scriptsize \left[ \begin{array}{c}
                      -y \\ x^2 \end{array} \right] }}{\longrightarrow} \:
{\cal O}^2_{ {\bf C}^2 } \:
\stackrel{ (x^2,y) }{\longrightarrow} \: {\cal O} \:
\longrightarrow \: D_x \: \longrightarrow \: 0
\end{displaymath}
to calculate local $\underline{\mbox{Ext}}^n_{ {\bf C}^2 }(D_x,2C)$.
Since the maps in the resolution vanish on the support of
$2C$, calculating local Ext is now trivial:
\begin{displaymath}
\underline{\mbox{Ext}}^n_{ {\cal O}_{ {\bf C}^2} }\left( D_x,
2C \right) \: = \:
\left\{ \begin{array}{cl}
        \underline{\mbox{Hom}}( {\cal O}, {\cal O}_0^2 ) \: = \:
{\cal O}_0^2 & n=0 \\
        \underline{\mbox{Hom}}( {\cal O}^2, {\cal O}_0^2 ) \: = \:
{\cal O}_0^4 & n=1 \\
        \underline{\mbox{Hom}}( {\cal O}, {\cal O}_0^2 ) \: = \:
{\cal O}_0^2 & n=2
        \end{array} \right.
\end{displaymath}
Applying the local-to-global spectral sequence, which is trivial
since the supports are on a point, 
we immediately recover the Serre dual of the Ext groups listed above,
as expected.

\subsection{Higgs vevs on only one side:  $(F, C)$}  \label{fc}

Consider open strings between three D0 branes at the origin of ${\bf C}^2$,
and a single D0 brane at the origin of ${\bf C}^2$.
Let the three D0 branes have nonzero Higgs vev, given by
\begin{displaymath}
\Phi^x \: = \: \left[ \begin{array}{ccc}
                      0 & 1 & 0 \\
                      0 & 0 & 0 \\
                      0 & 0 & 0 
                      \end{array} \right], \: \:
\Phi^y \: = \: \left[ \begin{array}{ccc}
                      0 & 0 & 1 \\
                      0 & 0 & 0 \\
                      0 & 0 & 0
                      \end{array} \right]
\end{displaymath}
Let us compute the open string spectra, using the BRST operator
$\overline{\partial} + \Phi^i_1 \theta_i - \Phi^i_2\theta_i$.
In this case, the $\Phi^i_2$ vanish, since there are nonzero Higgs vevs
on only one side of the open string.

First, let us compute the degree zero states.
Such states have the form
\begin{displaymath}
V \: = \: \left[ \begin{array}{c}
                  a \\ b \\ c
                  \end{array} \right]
\end{displaymath}
Here, $Q \cdot V = \Phi^x V \theta_1 + \Phi^y V \theta_2$,
so after a quick computation we find that the kernel of $Q$
has the form
\begin{displaymath}
\left[ \begin{array}{c}
       a \\ 0 \\ 0 \end{array} \right]
\end{displaymath}
so the space of physical states of degree zero has dimension one.

Next, consider the degree one states.
These have the form
\begin{displaymath}
V \: = \: \left[ \begin{array}{c}
                 a_x \\ b_x \\ c_x \end{array} \right] \theta_1
\: + \: \left[ \begin{array}{c} 
               a_y \\ b_y \\ c_y \end{array} \right] \theta_2 \: = \:
V_x \theta_1 \: + \: V_y \theta_2
\end{displaymath}
The BRST operator acts on $V$ as
\begin{displaymath}
Q \cdot V \: = \: \Phi^x V^y \theta_1 \theta_2 \: + \: \Phi^y V^x \theta_2
\theta_1 
\end{displaymath}
from which we can quickly compute that the kernel of $Q$ has the form
\begin{displaymath}
\left[ \begin{array}{c} 
       a_x \\ b_x \\ c_x \end{array} \right] \theta_1
\: + \:
\left[ \begin{array}{c}
       a_y \\ c_x \\ c_y \end{array} \right] \theta_2
\end{displaymath}
so we see that the kernel of $Q$, on degree one states,
has dimension five.
The image of $Q$ has the form
\begin{displaymath}
\left[ \begin{array}{c}
       b \\ 0 \\ 0 \end{array} \right] \theta_1 \: + \:
\left[ \begin{array}{c}
       c \\ 0 \\ 0 \end{array} \right] \theta_2
\end{displaymath}
for some $b$, $c$, and so has dimension two.
Since the kernel of $Q$ has dimension five, and the image of $Q$ has
dimension two, the dimension of the space of BRST-closed states
of degree one, modulo the BRST-exact states, is three.

Finally, consider the degree two states, of the form
\begin{displaymath}
V \: = \: \left[ \begin{array}{c}
                 a \\ b \\ c \end{array} \right] \theta_1 \theta_2
\end{displaymath}
$Q$ annihilates all of these states trivially, and the image of
degree one states under $Q$ has the form
\begin{displaymath}
\left[ \begin{array}{c}
       \alpha \\ 0 \\ 0 \end{array} \right]
\end{displaymath}
for some $\alpha$.  Hence the kernel has dimension three,
and the image has dimension one,
so the dimension of the space of BRST-closed states, modulo BRST-exact
states, is two.

In our conventions, these Higgs fields yield the structure sheaf
of the nonreduced scheme associated to the ideal $(x^2,xy,y^2)$,
which we are denoting by $F$.
Checking that $F$ is the output of these Higgs vevs is straightforward.
The element $x \in {\bf C}[x,y]$ acts as
\begin{displaymath}
\left[ \begin{array}{c}
       \alpha \\ \beta \\ \gamma \end{array} \right] 
\: \mapsto \:
\left[ \begin{array}{ccc}
       0 & 1 & 0 \\ 0 & 0 & 0 \\ 0 & 0 & 0 \end{array} \right]
\left[  \begin{array}{c}
       \alpha \\ \beta \\ \gamma \end{array} \right]
\: = \:
\left[ \begin{array}{c}
       \beta \\ 0 \\ 0 \end{array} \right]
\end{displaymath}
so $x$ maps the generator $\alpha$ to the generator $\beta$,
and annihilates both $\beta$ and $\gamma$.
Similarly, $y$ maps the generator $\alpha$ to the generator
$\gamma$, and annihilates both $\beta$ and $\gamma$.
If we identify $\alpha$ with the image of $1 \in {\bf C}[x,y]$
in the quotient ${\bf C}[x,y]/(x^2,xy,y^2)$,
$\beta$ with the image of $x$ in the quotient, and $\gamma$
with the image of $y$ in the quotient, then we see that the
module generated by the Higgs vevs in this case is precisely $F$.

Thus, our open string state computation predicts that
\begin{displaymath}
\mbox{dim }\mbox{Ext}^n_{ {\bf C}^2 } \left( F, C \right) \: = \:
\left\{ \begin{array}{cl}
        1 & n=0 \\
        3 & n=1 \\
        2 & n=2
        \end{array} \right.
\end{displaymath}
which is a true statement, and which we shall now check explicitly.

The sheaf $F$ has a locally-free resolution given by
\begin{displaymath}
0 \: \longrightarrow \:
{\cal O}_{ {\bf C}^2 }^2 \:
\stackrel{ {\scriptsize \left[ \begin{array}{cc}
           y & 0 \\ -x & y \\ 0 & -x \end{array} \right]}}{\longrightarrow}
\: {\cal O}_{ {\bf C}^2 }^3 \:
\stackrel{ \left[ x^2, xy, y^2 \right] }{\longrightarrow} \:
{\cal O}_{ {\bf C}^2 } \: \longrightarrow \: F \:
\longrightarrow \: 0
\end{displaymath}
so that the local $\underline{\mbox{Ext}}^*(F,C)$ sheaves are the cohomology
of the complex
\begin{displaymath}
0 \: \longrightarrow \:
\underline{\mbox{Hom}}\left({\cal O}, {\cal O}_0\right) \:
\longrightarrow \:
\underline{\mbox{Hom}}\left( {\cal O}^3, {\cal O}_0 \right) \:
\longrightarrow \:
\underline{\mbox{Hom}}\left( {\cal O}^2, {\cal O}_0 \right) \:
\longrightarrow \: 0
\end{displaymath}
Since the skyscraper sheaves $C = {\cal O}_0$ all have support where the
maps above all vanish, the cohomology of this complex is trivial to
compute, and so we find
\begin{displaymath}
\underline{\mbox{Ext}}^n_{ {\cal O}_{ {\bf C}^2 } }(F, C) \: = \:
\left\{ \begin{array}{cl}
        \underline{\mbox{Hom}}\left( {\cal O}, {\cal O}_0 \right) \: = \:
{\cal O}_0 & n=0 \\
        \underline{\mbox{Hom}}\left( {\cal O}^3, {\cal O}_0 \right) \: = \:
{\cal O}_0^3 & n=1 \\
        \underline{\mbox{Hom}}\left( {\cal O}^2, {\cal O}_0 \right) \: = \:
{\cal O}_0^2 & n=2 
        \end{array} \right.
\end{displaymath}
Applying the local-to-global spectra sequence, which is trivial
since the supports are on a point, we immediately recover the Ext groups
listed above, as expected.

Now, a careful reader might notice that if we replaced the Higgs
fields $\Phi^x$, $\Phi^y$ above with their transposes, and recomputed
open string spectra, we would get a different result.
Indeed -- the transposes of these Higgs fields describe a different
module, call it $\tilde{F}$, and repeating our physical analysis
of massless modes, we find that
\begin{displaymath}
\mbox{dim } \mbox{Ext}^n_{ {\bf C}^2 }\left( \tilde{F},C\right) \: = \:
\left\{ \begin{array}{cl}
        2 & n=0 \\
        3 & n=1 \\
        1 & n=2  \end{array} \right.
\end{displaymath}
This statement is easy to check.  The sheaf $\tilde{F}$ has
locally-free resolution
\begin{displaymath}
0 \: \longrightarrow \:
{\cal O}_{ {\bf C}^2 } \:
\stackrel{ {\scriptsize \left[ \begin{array}{c}
                               x^2 \\ xy \\ y^2 \end{array} \right] }
}{\longrightarrow} \: {\cal O}_{ {\bf C}^3 }^3 \:
\stackrel{ {\scriptsize \left[ \begin{array}{ccc}
                              y & -x & 0 \\
                              0 & y & -x \end{array} \right] }
}{\longrightarrow} \: {\cal O}_{ {\bf C}^2 }^2 \:
\longrightarrow \: \tilde{F} \: \longrightarrow \: 0
\end{displaymath}
It is straightforward to compute Ext groups, and we recover the same
result as the physical spectrum computation described above.

\subsection{Higgs vevs on both sides:  $(D_x, D_x)$}  \label{dxdx}

In this example we shall turn on Higgs fields on both sides
of the open strings, which run between two pairs of D0 branes
at the origin of ${\bf C}^2$.  We shall use the {\it same} Higgs fields
on both sides, given by
\begin{displaymath}
\Phi^x \: = \: \left[ \begin{array}{cc}
                      0 & 1 \\ 0 & 0  \end{array} \right], \: \:
\Phi^y \: = \: \left[ \begin{array}{cc}
                      0 & 0 \\ 0 & 0  \end{array} \right]
\end{displaymath}

Let us begin with the degree zero states.
These are of the form
\begin{displaymath}
V \: = \: \left[ \begin{array}{cc}
                 a & b \\ c & d  \end{array} \right]
\end{displaymath}
The BRST operator acts on such $V$ as
$Q \cdot V = \left( \Phi^x V - V \Phi^x \right) \theta_1$,
so it is straightforward to compute that the kernel of $Q$ is given
by 
\begin{displaymath}
\left[ \begin{array}{cc}
       a & b \\ 0 & a  \end{array} \right]
\end{displaymath}
and since this kernel is two-dimensional, and there is no image of $Q$ to
mod out, we see that the space of physical states of degree zero has
dimension two.

Next, consider the degree one states.
These can be written in the form
\begin{displaymath}
V \: = \: \left[ \begin{array}{cc}
                 a_x & b_x \\ c_x & d_x \end{array} \right] \theta_1
\: + \:
\left[ \begin{array}{cc}
       a_y & b_y \\ c_y & d_y  \end{array} \right] \theta_2 \: = \:
V^x \theta_1 \: + \: V^y \theta_2 
\end{displaymath}
The action of the BRST operator $Q$ on $V$ simplifies to
$Q \cdot V = \left( \Phi^x V^y - V^y \Phi^x \right) \theta_1 \theta_2$,
from which we can compute that the kernel of $Q$ has the form
\begin{displaymath}
\left[ \begin{array}{cc}
       a_x & b_x \\ c_x & d_x \end{array} \right] \theta_1 \: + \:
\left[ \begin{array}{cc}
       a_y & b_y \\ 0 & a_y \end{array} \right] \theta_2
\end{displaymath}
which is clearly six-dimensional.
The image of $Q$ has the form
\begin{displaymath}
\left[ \begin{array}{cc}
       a & b \\
       0 & a \end{array} \right] \theta_1
\end{displaymath}
and is clearly two-dimensional.
Thus, the space of physical states of degree one has dimension
$6-2=4$.

Finally, consider the degree two states,
which can be written in the form
\begin{displaymath}
V \: = \: \left[ \begin{array}{cc}
                 a & b \\ c & d  \end{array} \right] \theta_1 \theta_2
\end{displaymath}
All of these states are in the kernel of $Q$,
and the image of $Q$ has the form
\begin{displaymath}
\left[ \begin{array}{cc}
       \alpha & \beta \\ 0 & \alpha \end{array} \right] \theta_1 \theta_2
\end{displaymath}
for some $\alpha$, $\beta$, hence has dimension two.
Thus, the space of physical states of degree two has dimension $4-2=2$.

In section~\ref{2cdx} we argued that the Higgs fields used here on
both sides of the open string generate the sheaf $D_x$ from a pair
of skyscraper sheaves ${\cal O}_0^2 = 2C$.
Thus, in order for our mathematical ansatz to have physical
meaning, we need 
the massless spectrum computed above to match Ext groups between
the sheaves $D_x$; in other words:
\begin{displaymath}
\mbox{dim }\mbox{Ext}^n_{ {\bf C}^2 } \left( D_x, D_x \right) \: = \:
\left\{ \begin{array}{cl}
        2 & n=0 \\
        4 & n=1 \\
        2 & n=0
        \end{array} \right.
\end{displaymath}
This statement is true, and straightforward to check,
as we shall now outline.
Use the locally-free resolution of $D_x$ described in
section~\ref{2cdx}, the fact that
\begin{displaymath}
\underline{\mbox{Hom}}\left( {\cal O}, D_x \right) \: = \: D_x,
\end{displaymath},
and the fact that the maps in the induced complex all vanish,
then one has that
\begin{displaymath}
\underline{\mbox{Ext}}^n_{ {\cal O}_{ {\bf C}^2 } } \left( D_x, D_x \right)
\: = \:
\left\{ \begin{array}{cl}
        D_x & n=0 \\
        D_x^2 & n=1 \\
        D_x & n=2 
        \end{array} \right.
\end{displaymath}
The local-to-global spectral sequence degenerates, since these
sheaves have cohomology only in degree zero, and using the
fact that 
\begin{displaymath}
H^0\left( {\bf C}^2, D_x \right) \: = \: {\bf C}^2
\end{displaymath}
we recover the Ext groups above, exactly right to match the
physical spectrum calculation.

\subsection{Higgs vevs on both sides:  $(D_x, D_y)$}  \label{dxdy}

In this example we shall turn on Higgs fields on both sides of the
open strings, which run between two pairs of D0 branes at the origin
of ${\bf C}^2$.  We shall use different Higgs fields on the two sides,
given by
\begin{displaymath}
\Phi^x_1 \: = \: \left[ \begin{array}{cc}
                            0 & 1 \\ 0 & 0  \end{array} \right], \: \:
\Phi^y_1 \: = \: \left[ \begin{array}{cc}
                            0 & 0 \\ 0 & 0 \end{array} \right]
\end{displaymath}
which in sheaf language corresponds to the nonreduced scheme
$D_x$ as explained earlier,
and
\begin{displaymath}
\Phi^x_2 \: = \: \left[ \begin{array}{cc}
                            0 & 0 \\ 0 & 0  \end{array} \right], \: \:
\Phi^y_2 \: = \: \left[ \begin{array}{cc}
                            0 & 1 \\ 0 & 0   \end{array} \right]
\end{displaymath}
which in sheaf language corresponds to the nonreduced scheme $D_y$
 -- almost just as for $D_x$, except with a different scheme structure,
corresponding to two points colliding from different directions.

First, consider the degree zero states, which are of the form
\begin{displaymath}
V \: = \: \left[ \begin{array}{cc}
                 a & b \\ c & d  \end{array} \right]
\end{displaymath}
The action of the BRST operator $Q$ simplifies to
$Q \cdot V = \Phi^x_1 V \theta_1 - V \Phi^y_2 \theta_2$,
and it is straightforward to compute that the kernel has the form
\begin{displaymath}
\left[ \begin{array}{cc}
       0 & b \\ 0 & 0 \end{array}  \right]
\end{displaymath}
Since there is no image of $Q$ to mod out, we see that the space
of degree zero states has dimension one.

Next, consider the degree one states, which are of the form
\begin{displaymath}
V \: = \: \left[ \begin{array}{cc}
                 a_x & b_x \\ c_x & d_x \end{array} \right] \theta_1 \: + \:
\left[ \begin{array}{cc}
       a_y & b_y \\ c_y & d_y   \end{array} \right]
\: = \: V^x \theta_1 \: + \: V^y \theta_2
\end{displaymath}
The action of the BRST operator $Q$ simplifies to the form
$Q \cdot V = \Phi^x_1 V^y \theta_1 \theta_2 - V^x \Phi^y_2 \theta_2
\theta_1$
so the kernel of $Q$ has the form
\begin{displaymath}
\left[ \begin{array}{cc}
       a_x & b_x \\ 0 & d_x \end{array} \right] \theta_1 \: + \:
\left[ \begin{array}{cc}
       a_y & b_y \\ 0 & - a_x \end{array} \right] \theta_2
\end{displaymath}
and has dimension five.
The image of $Q$ has the form
\begin{displaymath}
\left[ \begin{array}{cc}
       c & d \\ 0 & 0  \end{array} \right] \theta_1 \: - \:
\left[ \begin{array}{cc}
       0 & a \\ 0 & c  \end{array} \right] \theta_2
\end{displaymath}
and has dimension three.
The space of physical states of degree one must therefore have
dimension $5-3=2$.

Finally, consider the degree two states, which are of the form
\begin{displaymath}
V \: = \: \left[ \begin{array}{cc}
                 a & b \\ c & d   \end{array} \right] \theta_1 \theta_2
\end{displaymath}
All of these states are in the kernel of $Q$,
and the image of $Q$ has the form
\begin{displaymath}
\left[ \begin{array}{cc}
       \alpha & \beta \\ 0 & \gamma \end{array} \right] \theta_1 \theta_2
\end{displaymath}
which has dimension three.
Thus, the space of physical states of degree two has dimension
$4-3=1$.

Thus, our analysis predicts
\begin{displaymath}
\mbox{dim }\mbox{Ext}^*_{ {\bf C}^2 } \left( D_x, D_y \right) \: = \:
\left\{ \begin{array}{cl}
        1 & n=0 \\
        2 & n=1 \\
        1 & n=2
        \end{array} \right.
\end{displaymath}
This statement is straightforward to check, as we shall now outline.
Use the locally-free resolution of $D_x$ given in section~\ref{2cdx},
and proceeding as in section~\ref{dxdx}, we find that the local
$\underline{\mbox{Ext}}$ sheaves are the cohomology of the complex
\begin{displaymath}
0 \: \longrightarrow \:
\underline{\mbox{Hom}}\left( {\cal O}, D_y \right) \:
\longrightarrow \:
\underline{\mbox{Hom}}\left( {\cal O}^2, D_y \right) \:
\longrightarrow \:
\underline{\mbox{Hom}}\left( {\cal O}, D_y \right) \:
\longrightarrow \: 0
\end{displaymath}
Also, $\underline{\mbox{Hom}}\left( {\cal O}, D_y \right) = D_y$.
However, unlike the closely analogous computation in section~\ref{dxdx},
here not all of the maps vanish.
For example, the map $D_y^2 \rightarrow D_y$ above proceeds
by composing local sections $(f,g)$ of the sheaf
\begin{displaymath}
\underline{\mbox{Hom}}\left( {\cal O}^2, D_y \right)
\: = \: D_y^2
\end{displaymath}
with $(x^2,y)$.  Now, such a composition would annihilate any section
of $D_x^2$, but here by contrast, although composition with $x^2$ annihilates
$f$, composition with $y$ does not annihilate $g$.  
Thus, one must be more careful when computing the cohomology of the complex.
Proceeding in this fashion one finds
\begin{displaymath}
\underline{\mbox{Ext}}^n_{ {\cal O}_{ {\bf C}^2 } }\left( D_x, D_y \right) 
\: = \:
\left\{ \begin{array}{cl}
        {\cal O}_0 & n=0 \\
        {\cal O}_0^2 & n=1 \\
        {\cal O}_0 & n=2
        \end{array} \right.
\end{displaymath} 
Applying the local-to-global spectral sequence as usual,
we recover the Ext groups above, precisely right to match the
physical spectrum calculation.

\subsection{Higgs vevs on both sides:  $(F,F)$}  \label{ff}

In this example we shall turn on Higgs vevs on both sides of the open strings,
which run between two sets of three D0 branes at the origin of ${\bf C}^2$.
We shall use the same set of Higgs fields on the two sides, given by
\begin{displaymath}
\Phi^x \: = \: \left[ \begin{array}{ccc}
                      0 & 1 & 0 \\
                      0 & 0 & 0 \\
                      0 & 0 & 0   \end{array}  \right], \: \:
\Phi^y \: = \: \left[ \begin{array}{ccc}
                      0 & 0 & 1 \\
                      0 & 0 & 0 \\
                      0 & 0 & 0    \end{array} \right]
\end{displaymath}
which correspond to the length three nonreduced scheme $F$, as
explained earlier.

First, consider degree zero states, which are of the form
\begin{displaymath}
V \: = \: \left[ \begin{array}{ccc}
                 a_{11} & a_{12} & a_{13} \\
                 a_{21} & a_{22} & a_{23} \\
                 a_{31} & a_{32} & a_{33}  \end{array} \right]
\end{displaymath}
The action of the BRST operator $Q$ on $V$ is summarized by
$Q \cdot V = \left( \Phi^x V - V \Phi^x \right) \theta_1 +
\left( \Phi^y V - V \Phi^y \right) \theta_2$,
{\it i.e.} to be in the kernel of $Q$,
$V$ must satisfy two equations:
\begin{eqnarray*}
\Phi^x V & = & V \Phi^x \\
\Phi^y V & = & V \Phi^y
\end{eqnarray*}
It is straightforward to check that $V$ satisfying these equations
can be written in the form
\begin{displaymath}
\left[ \begin{array}{ccc}
       a_{11} & a_{12} & a_{13} \\
          0 &   a_{11} & 0 \\
          0 & 0 &     a_{11}  \end{array} \right]
\end{displaymath}
Thus, since the kernel has dimension three and there is no image to mod out,
we see that the space of physical states of degree zero has dimension three.

Next, consider degree one states, which are of the form
\begin{displaymath}
V \: = \: \left[ \begin{array}{ccc}
                 a^x_{11} & a^x_{12} & a^x_{13} \\
                 a^x_{21} & a^x_{22} & a^x_{23} \\
                 a^x_{31} & a^x_{32} & a^x_{33} \end{array} \right] \theta_1
\: + \:
\left[ \begin{array}{ccc}
       a^y_{11} & a^y_{12} & a^y_{13} \\
       a^y_{21} & a^y_{22} & a^y_{23} \\
       a^y_{31} & a^y_{32} & a^y_{33}  \end{array} \right] \theta_2
\: = \:
V^x \theta_1 \: + \: V^y \theta_2
\end{displaymath}
The action of the BRST operator $Q$ on $V$ can be summarized as
$Q \cdot V = \Phi^x V^y \theta_1 \theta_2 + \Phi^y V^x \theta_2 \theta_1
- V^x \Phi_y \theta_2 \theta_1 - V^y \Phi_x \theta_1 \theta_2$,
and it is straightforward to check that states in the kernel of $Q$
have the form
\begin{displaymath}
\left[ \begin{array}{ccc}
       a^x_{11} & a^x_{12} & a^x_{13} \\
          0     & a^x_{22} & a^x_{23} \\
          0 & \left( a^y_{22} - a^y_{11} \right) & a^x_{33} \end{array}
         \right] \theta_1
\: + \:
\left[ \begin{array}{ccc}
       a^y_{11} & a^y_{12} & a^y_{13} \\
       0 & a^y_{22} & \left( a^x_{33} - a^x_{11} \right) \\
       0 & a^y_{32} & a^y_{33}  \end{array} \right] \theta_2
\end{displaymath}
which has dimension twelve.
The image of $Q$ in degree one states has the form
\begin{displaymath}
\left[ \begin{array}{ccc}
       a_{21} & \left( a_{22} - a_{11} \right) & a_{23} \\
       0 & - a_{21} & 0 \\
       0 & - a_{31} & 0  \end{array} \right] \theta_1
\: + \:
\left[ \begin{array}{ccc}
       a_{31} & a_{32} & \left( a_{33} - a_{11} \right) \\
       0 & 0 & -a_{21} \\
       0 & 0 & -a_{31}  \end{array} \right] \theta_2
\end{displaymath}
which has dimension six.
The space of physical states of degree one therefore has dimension
$12-6=6$.

Next, consider the degree two states, which are of the form
\begin{displaymath}
V \: = \: \left[ \begin{array}{ccc}
                 a_{11} & a_{12} & a_{13} \\
                 a_{21} & a_{22} & a_{23} \\
                 a_{31} & a_{32} & a_{33}  \end{array} \right]
\theta_1 \theta_2
\end{displaymath}
All of these states are annihilated by $Q$, and the image of $Q$ in
degree two states has the form
\begin{displaymath}
\left[ \begin{array}{ccc}
\left( a^y_{21} - a^x_{31} \right) & \left( a^y_{22} - a^x_{32} - a^y_{11}
\right) & \left( a^y_{23} - a^x_{33} + a^x_{11} \right) \\
0 & -a^y_{21} & a^x_{21} \\
0 & -a^y_{31} & a^x_{31}
\end{array} \right] \theta_1 \theta_2
\end{displaymath}
which has dimension six.
Thus, the space of physical states of degree two has dimension
$9-6=3$.

To be consistent with our physical computations above, we need that
\begin{displaymath}
\mbox{dim }\mbox{Ext}^n_{ {\bf C}^2 } \left( F, F \right) \: = \:
\left\{ \begin{array}{cl}
        3 & n=0 \\
        6 & n=1 \\
        3 & n=2
        \end{array} \right.
\end{displaymath}
and, again, this turns out to be true.
The calculation involved is almost identical to that
described in sections~\ref{dxdx} and \ref{dxdy}.
Here, using the locally-free resolution of $F$ described
in section~\ref{fc}, one quickly computes that the
local $\underline{\mbox{Ext}}$ sheaves are given by
the cohomology of the complex
\begin{displaymath}
0 \: \longrightarrow \: F \: \longrightarrow \: F^3 \: \longrightarrow
\: F^2 \: \longrightarrow \: 0
\end{displaymath}
The map $F \rightarrow F^3$ involves composing by $(x^2,xy,y^2)$,
which annihilates all sections, hence this map is identically zero.
The second map is not identically zero, and has the effect of
removing one $F$'s worth of sections from the local $\underline{\mbox{Ext}}$
sheaves.  Applying local-to-global as usual, one rapidly recovers
the Ext groups described above, precisely right to match the physical
spectrum computation.

\section{Example:  obstructed ${\bf P}^1$, or why normal bundles are not
local neighborhoods}  \label{obs}

A very interesting example involves D-branes wrapped on an obstructed
${\bf P}^1$ in a Calabi-Yau threefold.  Turning on a Higgs vev
corresponds to moving the sheaf inside the normal bundle,
but in this case, the curve cannot be moved inside the Calabi-Yau.
Since people typically identify Higgs vevs with finite deformations
inside the ambient space, rather than infinitesmial deformations /
deformations only inside the normal bundle, this example is both
interesting and important for both physical and mathematical reasons.

We will consider a single D-branes wrapped
on an obstructed ${\bf P}^1$, whose normal bundle
in the ambient Calabi-Yau threefold is ${\cal O} \oplus {\cal O}(-2)$.
The restriction of the tangent bundle $TX|_S$ does split holomorphically
as $TS \oplus {\cal N}_{S/X}$, so the usual notion of Higgs field
is applicable, and furthermore we shall assume the D-brane gauge bundle
is trivial, so as to simplify boundary conditions on worldsheet fermions.

Since the normal bundle contains an ${\cal O}$ factor, there is an
infinitesimal deformation of the D-brane -- a Higgs field, in the
sense of algebraic geometry -- corresponding to the holomorphic section
of the normal bundle.  But, how should that Higgs field be interpreted?

A better question is perhaps, if the curve is obstructed,
then how can the normal bundle be ${\cal O} \oplus {\cal O}(-2)$?
The total space of that bundle admits a one-parameter family of
${\bf P}^1$'s, including the zero section of the bundle (the original
${\bf P}^1$), yet we described the ${\bf P}^1$ as obstructed,
not admitting any finite deformations.

The mathematical resolution of this puzzle lies in the fact that in algebraic
geometry, normal bundles do {\it not} give a good local description
of the holomorphic geometry of the ambient space, unlike topology
and differential geometry, where the notion of a `tubular neighborhood'
is commonly used.  Instead, the holomorphic structure on the normal
bundle is only a linearization of the holomorphic structure on the
ambient space, and in this case, that linearization fails to capture
information about the obstruction.

More concretely, letting coordinates in two coordinate
patches be labelled by $(x, y_1, y_2)$ and $(w, z_1, z_2)$,
the Calabi-Yau complex structure locally about an obstructed ${\bf P}^1$ with
normal bundle ${\cal O} \oplus {\cal O}(-2)$ is defined by overlap maps
\begin{eqnarray*}
w & = & x^{-1} \\
z_1 & = & x^2 y_1 \: + \: x y_2^n \\
z_2 & = & y_2 
\end{eqnarray*}
(where $n$ is the degree of the obstruction)
whereas the complex structure on the total space of the normal
bundle ${\cal O} \oplus {\cal O}(-2)$ has overlaps
\begin{eqnarray*}
w & = & x^{-1} \\
z_1 & = & x^2 y_1 \\
z_2 & = & y_2 
\end{eqnarray*}
The ${\bf P}^1$ can be described in the respective coordinate neighborhoods
by the equations $y_1=y_2=0$ (resp.\ $z_2=z_2=0$) in both the Calabi-Yau 
and in the normal bundle.\footnote{In the case of the normal bundle,
this ${\bf P}^1$ is identified with the zero section.}  The ${\bf P}^1$ can
be deformed by keeping the equations $y_1=z_1=0$ and deforming $y_2=z_2=0$
to $y_2=z_2=\epsilon$.  This deformation is valid for all $\epsilon$ in 
the normal bundle, but in the Calabi-Yau, consistency with $z_1=0$
forces the obstruction $\epsilon^n=0$.

This should make it explicitly clear that the complex structure
on the normal bundle is only a linearization of the complex structure
on the ambient space, a linearization that omits the obstruction data.
In particular, although the ${\bf P}^1$
can be deformed inside the normal bundle, it cannot be deformed
inside the ambient space.

More generally this issue that holomorphically the total space
of the normal bundle is not a good model of the complex structure
of the ambient space is very much a headache for algebraic geometry.
For example, a substantial portion of \cite{fulton} is devoted to workarounds
for this issue.

Physically, in this case although the operator corresponding to the
Higgs field is marginal, it is not truly marginal, as some higher
correlation functions do not vanish,
but rather encode the
mathematical obstruction data.  Giving a vev to this Higgs field
breaks conformal invariance, albeit in a subtle fashion.
See \cite{merev} for more information on the physics of D-branes
wrapped on the obstructed ${\bf P}^1$.

The reader should now be better equipped to understand remarks
made earlier, that giving a vev to a Higgs field should be
interpreted as moving the D-brane inside its normal bundle,
but not the ambient space.  Here, giving a vev to the Higgs field
creates a new sheaf inside the total space of the normal bundle
whose support has been shifted away from the original ${\bf P}^1$,
completely consistent with the fact that
the ${\bf P}^1$ cannot be moved inside the ambient space.

\section{Commutativity of Higgs vevs versus unobstructedness of moduli}
\label{fterms}

Earlier we mentioned that in order to consistently give a Higgs
field a vev, one constraint (arising from demanding $Q^2=0$) is that
the Higgs vevs must commute with one another:  $[\Phi^i, \Phi^j] = 0$,
a condition that in the target space theory is typically an F-term
condition.

Mathematically this condition is a necessary, but not sufficient,
condition for a modulus to be unobstructed.  We will take a moment
to review this fact.

In general, given an infinitesimal modulus of a sheaf $i_* {\cal E}$,
corresponding to an element of
\begin{displaymath}
\mbox{Ext}^1_X\left( i_* {\cal E}, i_* {\cal E} \right)
\end{displaymath}
the first obstruction to deforming a finite distance in that direction
is given by the Yoneda pairing:
\begin{displaymath}
\mbox{Ext}^1_X\left( i_* {\cal E}, i_* {\cal E} \right)
\: \times \:
\mbox{Ext}^1_X\left( i_* {\cal E}, i_* {\cal E} \right)
\: \longrightarrow \:
\mbox{Ext}^2_X\left( i_* {\cal E}, i_* {\cal E} \right).
\end{displaymath}

The Yoneda pairing fits into a commutative diagram:
\begin{displaymath}
\xymatrix{
\mbox{Ext}^1_X\left( i_* {\cal E}, i_* {\cal E} \right) \ar[d]
& \times &
\mbox{Ext}^1_X\left( i_* {\cal E}, i_* {\cal E} \right)
 \ar[r] \ar[d] & 
\mbox{Ext}^2_X\left( i_* {\cal E}, i_* {\cal E} \right) \ar[d] \\
H^0\left(S, {\cal E}^{\vee} \otimes {\cal E} \otimes {\cal N}_{S/X} \right)
& \times &
H^0\left(S, {\cal E}^{\vee} \otimes {\cal E} \otimes {\cal N}_{S/X} \right)
\ar[r] &
H^0\left(S, {\cal E}^{\vee} \otimes {\cal E} \otimes \Lambda^2{\cal N}_{S/X}
\right)
}
\end{displaymath}
and the commutivity statement is just the statement that
the image of the bottom product in 
$H^0\left(S, {\cal E}^{\vee} \otimes {\cal E} \otimes \Lambda^2{\cal N}_{S/X}
\right)$ vanishes.
(The necessary condition for the modulus to be unobstructed is that
the image in $H^0\left(S, {\cal E}^{\vee} \otimes {\cal E} \otimes \Lambda^2{\cal N}_{S/X}
\right)$ vanishes in cohomology, but since this is a degree {\it zero}
cohomology class, for this to vanish in cohomology means that
the wedge product must vanish identically.)

\section{Orbifolds}   \label{orbs}

One of the motivations for this work is to help clarify
the physical meaning of results in \cite{dgm} for
D-branes on orbifolds.  There it was found that the classical
Higgs moduli space of the low-energy gauge theory of D-branes
in orbifolds is a resolution of the quotient space,
a result which has led some physicists to claim that string
orbifolds describe strings on resolutions of quotient spaces.

Re-reading \cite{dgm} reveals that the classical Higgs moduli spaces
encode resolution of quotient spaces because D-terms no longer
exclude nilpotent Higgs vevs, which are responsible for the
exceptional divisors.  

{}From our previous examples, we have seen that nilpotent Higgs vevs
typically lead to {\it e.g.} structure sheaves of nonreduced schemes,
and indeed, it has been argued (see {\it e.g.} \cite{mohri}) that
the classical Higgs moduli spaces seen by D-branes are exactly
moduli spaces of nonreduced schemes ({\it i.e.} Hilbert schemes).
Thus, our methods give a new understanding of that old result.

Furthermore, the McKay correspondence in the form \cite{bkr}
typically gives nonreduced schemes in places where one would
expect nilpotent Higgs fields.  Our results help illustrate
the consistency of \cite{bkr} with calculations in D-branes.

Our results also shed light on how the idea that string orbifolds
are strings on quotient stacks \cite{qs} can be compatible with
the calculations described above.  We see that the nilpotent Higgs vevs,
responsible for the exceptional divisors in classical Higgs moduli spaces,
are equivalent to $G$-equivariant nonreduced schemes on the covering
space, {\it i.e.} nonreduced schemes on the quotient stack.
Thus, it is consistent both for classical Higgs moduli spaces
of D-branes in orbifolds to see resolutions of quotient spaces,
while simultaneously the strings themselves propagate on quotient stacks,
and indeed, this was the mechanism (using \cite{tomasme}) used in
\cite{qs} to explain this apparent discrepancy.

In this section we shall further illuminate these matters,
by showing in examples precisely how the exceptional divisors
in classical Higgs moduli spaces of D-branes on orbifolds are
nonreduced schemes, and by further describing how this is
consistent with the McKay correspondence in the form \cite{bkr}.

\subsection{Exceptional divisors are nilpotent Higgs fields}
\label{dbraneorbcomp}

In \cite{dgm} it was observed that the classical moduli space
of D-branes on an orbifold $[X/G]$ is not the quotient space
$X/G$, but rather a {\it resolution} $\widetilde{X/G}$ of the quotient space.
This observation has been the basis for {\it e.g.} interpretations of
string orbifolds as strings on resolutions of quotient spaces.
However, those exceptional divisors are, strictly speaking,
arising from nilpotent Higgs vevs, as we shall now review.

This can be seen in general as follows.
Demanding that the Higgs fields commute with the orbifold group
action, together with the F-term conditions which imply
that the Higgs fields commute with one another,
implies that the Higgs fields are simultaneously diagonalizable
with eigenvalues/eigenvectors related by the orbifold group action.
We can project onto the quotient space by mapping a given 
set of Higgs fields to its eigenvalues -- since the eigenvalues
are related by the orbifold group action on the cover,
this gives a well-defined point on the quotient space.
This map fails to be one-to-one precisely when there are
nilpotent components, which can only happen over fixed points
of the orbifold group action.  The effect is as observed in
\cite{dgm} -- because of nilpotent Higgs fields, the classical
Higgs moduli space is a resolution of the quotient space.

To make this more explicit,
let us review in detail the classical moduli space of D0 branes
on  $[ {\bf C}^2/{\bf Z}_2]$, repeating the analysis
of \cite{dgm}.  We will see explicitly that
the exceptional divisor of the resolution, as see by the classical
moduli space of the D-branes, corresponds to nilpotent Higgs fields.

For simplicity, we shall consider two D0 branes, both supported
at the origin of ${\bf C}^2$, hence described by the direct
sum of two skyscraper sheaves at the origin.

In order to define the ${\bf Z}_2$ orbifold, we need to choose
a ${\bf Z}_2$-action on the Chan-Paton factors,
{\it i.e.},
a ${\bf Z}_2$-equivariant structure on the corresponding
pair of skyscraper sheaves.
In such a simple case, the equivariant structure is 
really just a choice of two-dimensional representation
of ${\bf Z}_2$.  We shall use the regular representation.

The fields on the D0 branes in the orbifold are ${\bf Z}_2$-invariants.
Thus, given 2 D0 branes as above, we have two fields, call them
$X$ and $Y$, which are the Higgs fields on the D0 brane worldvolume,
and arise from open strings connecting the D0 branes to themselves.
$X$ and $Y$ are both $2 \times 2$ matrices.
In the orbifold theory, we take the ${\bf Z}_2$ invariants,
and it's easy to check that that means $X$ and $Y$ must have the form
\begin{displaymath}
X \: = \: \left[ \begin{array}{cc}
                 0 & x_1 \\
                 x_0 & 0 \end{array} \right], \:
Y \: = \: \left[ \begin{array}{cc}
                 0 & y_1 \\
                 y_0 & 0  \end{array} \right]
\end{displaymath}
We can work in a complexified theory and think of $x_{0,1}$, $y_{0,1}$
as all being complex numbers.

The original $U(2)$ 
gauge symmetry is reduced, by the orbifold projection,
to $U(1)^2$.
The matrices above transform as adjoints under this
gauge symmetry.  It is easy to check that one of the $U(1)$'s decouples,
and under the other $U(1)$, $x_0$, $y_0$ have the same charge
(call it $+1$), and $x_1$, $y_1$ have the opposite charge
(call it $-1$).

So far we have described the Higgs fields on the D0 brane worldvolume,
arising from open strings connecting the D0 branes to themselves.

In order to describe classical vacua of the theory,
these fields must satisfy additional constraints\footnote{Strictly
speaking, these constraints emerge from a triplet of D-terms, given
the amount of supersymmetry present; however, we shall refer to them
as F- and D-term constraints, in reference to \cite{dgm}.}:
\begin{enumerate}
\item F term constraints:  $[X,Y] = 0$.
It is straightforward to check that the condition that the two
matrices commute reduces to the single equation $x_1 y_0 = x_0 y_1$.
\item D term constraints:
\begin{displaymath}
| x_0 |^2 \: + \: | y_0 |^2 \: - \: | x_1 |^2 \: - \: | y_1 |^2 \: = \: r
\end{displaymath}
where $r$ is a constant.
This is, of course, part of a symplectic quotient corresponding to
the nontrivial $U(1)$.
\end{enumerate}

After modding out the remaining $U(1)$, or,
if one prefers, performing a GIT quotient wherein
$(x_0, x_1, y_0, y_1)$ have weights $(+1, -1, +1, -1)$,
the classical moduli space of Higgs fields is given by
\begin{displaymath}
\{ ( x_1 y_0 = y_1 x_0) \subset {\bf C}^4 \} // {\bf C}^{\times}
\end{displaymath}
It can be shown\footnote{Describe $\widetilde{ {\bf C}^2 / {\bf Z}_2 }$ as
$ {\bf C}^3 // {\bf C}^{\times}$,
where if we label coordinates on ${\bf C}^3$ by $(x,y,p)$,
the coordinates have weights $(+1,+1,-2)$ under the ${\bf C}^{\times}$.
Then identify $x_0 = x$, $x_1=xp$, $y_0=y$, and $y_1=yp$.} 
that this quotient is the same as the minimal
resolution of the quotient space ${\bf C}^2 / {\bf Z}_2$.

We can also outline the result using basic linear algebra.
Since the original matrices $X$ and $Y$ commute, they're simultaneously
diagonalizable.  It is easy to check that since they're ${\bf Z}_2$
invariant, the (nonzero) eigenvalues come in ${\bf Z}_2$ pairs.
If we map the pair $(X,Y)$ to the point on $\widetilde{ {\bf C}^2/
{\bf Z}_2 }$ determined by the eigenvalues, then for nonzero
eigenvalues, we have an isomorphism.
The zero eigenvalues are degenerate -- they are mapping out the
exceptional divisor.

In this language, it is clear that the exceptional divisor
in $\widetilde{ {\bf C}^2 / {\bf Z}_2 }$ is arising from
nilpotent matrices $X$, $Y$.
In particular, if $r > 0$, then the matrices corresponding to
the exceptional divisor are given by
\begin{displaymath}
X \: = \: \left[ \begin{array}{cc}
                 0 & 0 \\
                 x_0 & 0  \end{array} \right], \:
Y \: = \: \left[ \begin{array}{cc}
                 0 & 0 \\
                 y_0 & 0 \end{array} \right]
\end{displaymath}
where $x_0$, $y_0$ are homogeneous coordinates on the ${\bf P}^1$.
The corresponding length 2 ideal on ${\bf C}[x,y]$ is given by
$(x^2, x_0 y -  y_0 x)$.
We have seen the corresponding sheaves previously in this text:
$D_x$ is the sheaf corresponding to the case $y_0=0$,
and $D_y$ is the sheaf corresponding to the case $x_0=0$.

We have only discussed the special case $[ {\bf C}^2/{\bf Z}_2]$,
but as should be clear from the linear-algebra-based discussion
above, the same ideas work in generality -- D-branes on orbifolds
see resolutions of quotient spaces because the gauge theory classical moduli
space admits nilpotent Higgs vevs.  
See \cite{dgm} for a much more general discussion.

\subsection{The McKay correspondence -- exceptional divisors are nonreduced
schemes}   \label{mckay}

In the previous section, we showed that the exceptional divisors
in moduli spaces of D-branes in orbifolds are determined by
nilpotent Higgs vevs.  There is another way to think about
D-brane moduli spaces, utilizing the McKay correspondence,
which generates nonreduced schemes, as we shall outline here.
Clearly, for consistency one would like nonreduced schemes
to correspond physically to D-branes with nilpotent Higgs vevs,
which is exactly what we have already observed.

The version of the McKay correspondence that is applicable
here is due to Bridgeland-King-Reid \cite{bkr}, who described
McKay at the level of an equivalence of derived categories of sheaves.
If one starts with a skyscraper sheaf on the exceptional divisor
of $\widetilde{ {\bf C}^2/{\bf Z}_2 }$,
then just as discussed in \cite[section 6.2]{kps} for skyscraper
sheaves on $\widetilde{ {\bf C}^3/{\bf Z}_3 }$,
the image under the McKay functor is a ${\bf Z}_2$-equivariant
nonreduced scheme on ${\bf C}^2$, with support at the origin
(the ${\bf Z}_2$ fixed point), and with scheme structure determined
by the location of the skyscraper sheaf on the exceptional divisor.

Clearly, to be consistent with \cite{dgm}, one would like
for nonreduced schemes to be related to nilpotent Higgs fields,
and that is exactly what we have found.

Also, in followup work to \cite{dgm}, it was noticed that the
resolutions could also be obtained as Hilbert schemes of points,
{\it i.e.} moduli spaces of $G$-equivariant nonreduced schemes
(see {\it e.g.} \cite{mohri}), as relevant for another version
of the McKay correspondence \cite{itonakajima}.
One interpretation of the present work is that we are giving
a detailed physical justification for that correspondence.

The McKay correspondence of \cite{bkr} preserves Ext groups,
so one can calculate Ext groups
between nonreduced schemes in orbifolds by relating them to 
Ext groups between corresponding skyscraper sheaves on resolutions.
For example, in section~\ref{dxdyz2} we shall calculate
Ext groups on the quotient stack $[{\bf C}^2/{\bf Z}_2]$
between nonreduced schemes $D_x$, $D_y$ corresponding to distinct
points on the exceptional divisor, and we shall see that
\begin{displaymath}
\mbox{Ext}^n_{ [ {\bf C}^2/{\bf Z}_2 ] } \left( D_x, D_y \right) \: = \:
0 \mbox{ for all }n, \mbox{ unless }x=y
\end{displaymath}
which is exactly what one would expect, as on the resolution of the
quotient space the corresponding D0-branes do not intersect.
In the case that $y=x$, so that the D0-branes on the resolution coincide,
we shall see in section~\ref{dxdxz2} that
\begin{displaymath}
\mbox{Ext}^n_{  [ {\bf C}^2/{\bf Z}_2 ] } \left( D_x, D_x \right) \: = \:
\left\{ \begin{array}{cl}
        {\bf C} & n=0 \\
        {\bf C}^2 & n=1 \\
        {\bf C} & n=2 \\
        0 & n>0
        \end{array} \right.
\end{displaymath}
which, just as one would hope, matches the spectrum of open strings
from a single D0-brane back to itself on a smooth two-dimensional space.

\section{Examples in orbifolds -- coincident D0 branes}  \label{orbexs}

In this section we shall calculate several examples of massless
boundary Ramond sector spectra between D-branes with Higgs vevs
in orbifolds, following \cite{kps}.  Massless spectra in orbifolds
are counted by Ext groups on quotient stacks, as shown in
\cite{kps}, so we will be checking that massless spectra between
D-branes with Higgs vevs are counted by Ext groups on quotient stacks,
between sheaves obtained by our mathematical ansatz.

Although this might sound slightly intimidating, the details
are very straightforward.  A sheaf on a quotient stack $[X/G]$
is precisely a $G$-equivariant sheaf on $X$, so it is easy to
apply our mathematical ansatz to construct sheaves on quotient stacks
 -- just keep track of the $G$ action on the Chan-Paton factors,
and use only Higgs fields that commute with the $G$ action.
Similarly, Ext groups on a quotient stack $[X/G]$, for finite $G$,
are the same as $G$-invariant parts of Ext groups on $X$ between
the sheaves.

Our notation is the same as before.  We use $C$ to denote a skyscraper
sheaf at the origin of ${\bf C}^2$ and $D_x$ to denote a sheaf obtained
from $2C$ with a nilpotent Higgs field.
We will consider ${\bf Z}_2$ orbifolds of ${\bf C}^2$.
The ${\bf Z}_2$ actions are defined as follows:  $2C$ has ${\bf Z}_2$
action defined by the regular representation of ${\bf Z}_2$,
and $D_x$ is obtained from $2C$ in the regular representation of
${\bf Z}_2$, by using the same nilpotent Higgs fields as before
that, not coincidentally, happen to commute with the ${\bf Z}_2$
action.

\subsection{No Higgs vevs either side:  $(2C, 2C)$}

Consider two pairs of D0 branes at the origin of ${\bf C}^2$, as in
section~\ref{2c2c}.  Consider the usual supersymmetric ${\bf Z}_2$ action
on the ${\bf C}^2$, and define the ${\bf Z}_2$ action on the Chan-Paton
factors by putting each pair of D0 branes in the regular representation
of ${\bf Z}_2$.  

Computing open string spectra in the present case -- with no Higgs vevs
-- is a special case of computations in \cite{kps}, but for completeness,
we shall review them here, and also set the stage for later examples.

Now, not all Higgs vevs are allowed -- they must commute with
the orbifold group action.  In this case, we must demand,
for each of 
\begin{displaymath}
\Phi^{x,y} \: = \: \left[ \begin{array}{cc}
                          a & b \\
                          c & d \end{array} \right]
\end{displaymath}
that
\begin{displaymath}
\left[ \begin{array}{cc}
       1 & 0 \\ 0 & -1  \end{array} \right]
\left[ \begin{array}{cc}
       a & b \\ c & d \end{array} \right]
\left[ \begin{array}{cc}
       1 & 0 \\
       0 & -1   \end{array} \right]^{-1} \: = \: 
- \left[ \begin{array}{cc}
         a & b \\ c & d \end{array} \right]
\end{displaymath}
which implies that $a=d=0$ but $b$, $c$ are arbitrary.
It is not a coincidence that the Higgs vevs we used in section~\ref{d0exs}
to define the schemes $D_x$, $D_y$ all have this property.

Now, to compute the spectrum of open strings on the orbifold
$[ {\bf C}^2 / {\bf Z}_2 ]$, we simply take ${\bf Z}_2$-invariants,
as described in \cite{kps}.

Degree zero states on the covering space are of the form
\begin{displaymath}
\left[ \begin{array}{cc}
       a & b \\
       c & d  \end{array} \right]
\end{displaymath}
The ${\bf Z}_2$-invariant states are those that obey
\begin{displaymath}
\left[ \begin{array}{cc}
       1 & 0 \\
       0 & -1 \end{array} \right]
\left[ \begin{array}{cc}
       a & b \\
       c & d \end{array} \right] 
\left[ \begin{array}{cc}
       1 & 0 \\
       0 & -1 \end{array} \right]^{-1}
\: = \:
\left[ \begin{array}{cc}
       a & b \\
       c & d  \end{array} \right]
\end{displaymath}
which implies $b=c=0$ but $a, d$ are arbitrary.
Thus, the space of physical states of degree zero is two dimensional.

Degree one states on the covering space are of the form
\begin{displaymath}
V \: = \:
\left[ \begin{array}{cc}
       a_1 & b_1 \\
       c_1 & d_1  \end{array} \right] \theta_1 \: + \:
\left[ \begin{array}{cc}
       a_2 & b_2 \\
       c_2 & d_2  \end{array} \right] \theta_2
\end{displaymath}
Since the ${\bf Z}_2$ acts nontrivially on the $\theta_i$, the condition
for the degree one states to be ${\bf Z}_2$-invariant is that
\begin{displaymath}
\left[ \begin{array}{cc}
       1 & 0 \\ 0 & -1  \end{array} \right]
\left[ \begin{array}{cc}
       a_i & b_i \\ c_i & d_i \end{array} \right]
\left[ \begin{array}{cc}
       1 & 0 \\
       0 & -1   \end{array} \right]^{-1} \: = \: 
- \left[ \begin{array}{cc}
         a_i & b_i \\ c_i & d_i \end{array} \right]
\end{displaymath}
for each $i$, so $a_i=d_i=0$ but $b_i$ and $c_i$ are arbitrary, for each $i$.
Thus, the space of physical states of degree one is four dimensional.

Degree two states on the covering space are of the form
\begin{displaymath}
\left[ \begin{array}{cc}
       a & b \\
       c & d  \end{array} \right]
\theta_1 \theta_2
\end{displaymath}
The ${\bf Z}_2$ acts by a sign on each $\theta$, and so leaves
the product $\theta_1 \theta_2$ invariant.  The analysis here is identical
to that for degree zero states, so clearly the space of physical
states of degree two is two dimensional.

Thus, in order to match the physical spectrum computation,
we require that
\begin{displaymath}
\mbox{dim } \mbox{Ext}^n_{ [ {\bf C}^2 / {\bf Z}_2 ] }
\left( 2C, 2C \right) \: = \:
\left\{ \begin{array}{cl}
        2 & n=0 \\
        4 & n=1 \\
        2 & n=2
        \end{array} \right.
\end{displaymath}
which is a true statement.

\subsection{Higgs vevs on only one side:  $(2C, D_x)$ }

Next we shall consider two pairs of D0 branes at the origin of ${\bf C}^2$,
as before, but now with nonzero Higgs vevs on one side,
specifically,
\begin{displaymath}
\Phi^x \: = \: \left[ \begin{array}{cc}
                       0 & 1 \\
                       0 & 0
                       \end{array} \right],
\: \:
\Phi^y \: = \: \left[ \begin{array}{cc}
                      0 & 0 \\
                      0 & 0
                      \end{array} \right]
\end{displaymath}
Each pair of D0 branes is put in the regular representation of
${\bf Z}_2$, as before.
The Higgs vevs must commute with the orbifold group action, and it is easy
to verify that, indeed, these Higgs vevs do commute with the orbifold
group action.
 
As before, to compute open string states in orbifolds by finite groups,
we compute on the covering space and then take group invariants.

As computed in section~\ref{2cdx}, the degree zero states on the covering
space are of the form
\begin{displaymath}
\left[ \begin{array}{cc}
       0 & b \\
       0 & d 
       \end{array} \right]
\end{displaymath}
The orbifold group acts by conjugating these states by
\begin{displaymath}
\left[ \begin{array}{cc}
       1 & 0 \\ 0 & -1  \end{array} \right]
\end{displaymath}
which sends
\begin{displaymath}
\left[ \begin{array}{cc}
       0 & b \\ 0 & d \end{array} \right] \: \mapsto \:
\left[ \begin{array}{cc}
       0 & -b \\ 0 & d  \end{array} \right]
\end{displaymath}
so the only physical degree zero states that survive the orbifold projection
are 
\begin{displaymath}
\left[ \begin{array}{cc}
       0 & 0 \\
       0 & d   \end{array} \right]
\end{displaymath}
from which we see that the space of physical states
of degree zero has dimension one.

{}From section~\ref{2cdx} recall that the degree one states in the kernel of $Q$,
modulo the image of $Q$, could be written in the form
\begin{displaymath}
V \: = \:
\left[ \begin{array}{cc}
       a_x & 0 \\
       c_x & 0  \end{array} \right] \theta_1 \: + \:
\left[ \begin{array}{cc}
       0 & b_y \\
       0 & d_y  \end{array} \right] \theta_2
\end{displaymath}
The orbifold group acts by conjugating the matrices, just as for
degree zero states, but also multiplies each of the $\theta$'s by
$-1$.  Thus, the degree one physical states that survive the orbifold
projection are of the form
\begin{displaymath}
\left[ \begin{array}{cc}
0 & 0 \\
c_x & 0 \end{array} \right] \theta_1 \: + \:
\left[ \begin{array}{cc}
0 & b_y \\
0 & 0   \end{array} \right] \theta_2
\end{displaymath}
and so we see that the space of physical states of degree one has
dimension two.

Finally, as shown in section~\ref{2cdx},
the degree two states on the covering space, modulo the image of $Q$, have
the form
\begin{displaymath}
\left[ \begin{array}{cc}
       a & 0 \\
       c & 0 \end{array} \right] \theta_1 \theta_2
\end{displaymath}
The orbifold group acts on the degree two states in the same
way as the degree zero states, so we see that the states surviving
the orbifold projection have the form
\begin{displaymath}
\left[ \begin{array}{cc}
       a & 0 \\
       0 & 0  \end{array} \right] \theta_1 \theta_2
\end{displaymath}
Thus, the space of physical states of degree two in the orbifold theory
has dimension one.

Thus, in order for the physical spectrum computation
to match Ext group computations, we require
\begin{displaymath}
\mbox{dim } \mbox{Ext}^n_{ [ {\bf C}^2 / {\bf Z}_2 ] }
\left( 2C, D_x \right) \: = \:
\left\{ \begin{array}{cl}
        1 & n=0 \\
        2 & n=1 \\
        1 & n=2  \end{array} \right.
\end{displaymath}
which is a true statement.

\subsection{Higgs vevs on both sides:  $(D_x, D_x)$ }   \label{dxdxz2}

Next let us turn on Higgs vevs on both pairs of D0 branes in the
orbifold, the same Higgs vevs on each side in this example.
The Higgs vevs we shall consider are the same as in section~\ref{dxdx}:
\begin{displaymath}
\Phi^x \: = \: \left[ \begin{array}{cc}
                       0 & 1 \\
                       0 & 0
                       \end{array} \right],
\: \:
\Phi^y \: = \: \left[ \begin{array}{cc}
                      0 & 0 \\
                      0 & 0
                      \end{array} \right]
\end{displaymath}
As before, we shall assume ${\bf Z}_2$ action on the Chan-Paton factors is
defined by the regular representation; the Higgs vev above is consistent
with that representation, as discussed previously.

{}From section~\ref{dxdx}, recall the physical degree zero states 
on the covering space ${\bf C}^2$ are
of the form
\begin{displaymath}
\left[ \begin{array}{cc}
       a & b \\
       0 & a   \end{array} \right]
\end{displaymath}
The orbifold group acts by conjugating the matrix above by
$\mbox{diag}(1,-1)$, as before, so the physical states that are
invariant under the orbifold group action are of the form
\begin{displaymath}
\left[ \begin{array}{cc}
       a & 0 \\
       0 & a  \end{array} \right]
\end{displaymath}
Thus, the space of physical states of degree zero in the orbifold
theory has dimension one.

{}From section~\ref{dxdx}, recall the physical degree one states on the
covering space ${\bf C}^2$ have the form
\begin{displaymath}
\left[ \begin{array}{cc}
       0 & 0 \\
       c_x & d_x \end{array} \right] \theta_1 \: + \:
\left[ \begin{array}{cc}
       a_y & b_y \\
       0 & a_y \end{array} \right] \theta_2
\end{displaymath}
The orbifold group acts by conjugating each of the matrices
by $\mbox{diag}(1,-1)$, and by multiplying the $\theta$'s by $-1$.
Thus, the physical degree one states that are invariant under the
orbifold group action have the form
\begin{displaymath}
\left[ \begin{array}{cc}
       0 & 0 \\
       c_x & 0 \end{array} \right] \theta_1 \: + \:
\left[ \begin{array}{cc}
       0 & b_y \\
       0 & 0  \end{array} \right] \theta_2
\end{displaymath}
and this space has dimension two.

{}From section~\ref{dxdx}, recall the physical degree two states on the
covering space ${\bf C}^2$ have the form
\begin{displaymath}
\left[ \begin{array}{cc}
       0 & 0 \\
       c & d \end{array} \right] \theta_1 \theta_2
\end{displaymath}
The orbifold group acts by conjugating the matrix by $\mbox{diag}(1,-1)$,
and so the physical states that are invariant under the
orbifold projection have the form
\begin{displaymath}
\left[ \begin{array}{cc}
       0 & 0 \\
       0 & d  \end{array} \right] \theta_1 \theta_2
\end{displaymath}
which have dimension one.

Thus, in order for the physical spectrum computation to match
Ext groups, we require
\begin{displaymath}
\mbox{dim } \mbox{Ext}^n_{ [ {\bf C}^2 / {\bf Z}_2 ] }
\left( D_x, D_x \right) \: = \:
\left\{ \begin{array}{cl}
        1 & n=0 \\
        2 & n=1 \\
        1 & n=2
        \end{array} \right.
\end{displaymath}
which is a true statement, and matches results obtained
using the McKay correspondence in section~\ref{mckay}.

\subsection{Higgs vevs on both sides:  $(D_x, D_y)$}   \label{dxdyz2}

Next let us turn on Higgs vevs on both pairs of D0 branes in
the orbifold, but different Higgs vevs on either side, of the
same form as those in section~\ref{dxdy}.
As before, we shall assume the ${\bf Z}_2$ action on the Chan-Paton
factors is defined by the regular representation; the Higgs vevs
used in section~\ref{dxdy} are consistent with that representation,
as discussed previously.

{}From section~\ref{dxdy}, recall the physical degree zero states on
the covering space ${\bf C}^2$ are of the form
\begin{displaymath}
\left[ \begin{array}{cc}
       0 & b \\  0 & 0  \end{array} \right]
\end{displaymath}
Just as in the previous section, the orbifold group acts by
conjugating the matrix above by $\mbox{diag}(1,-1)$,
which acts nontrivially on $b$.  Thus, only the zero matrix is
${\bf Z}_2$-invariant, and so the space of physical states of
degree zero has dimension zero.

{}From section~\ref{dxdy}, recall the physical degree one states on
the covering space ${\bf C}^2$ are states of the form
\begin{displaymath}
\left[ \begin{array}{cc}
       a_x & b_x \\ 0 & d_x \end{array} \right] \theta_1
\: + \:
\left[ \begin{array}{cc}
       a_y & b_y \\ 0 & -a_x \end{array} \right] \theta_2
\end{displaymath}
modulo states of the form
\begin{displaymath}
\left[ \begin{array}{cc}
       c & d \\ 0 & 0  \end{array} \right] \theta_1 \: - \:
\left[ \begin{array}{cc}
       0 & a \\ 0 & c \end{array} \right] \theta_2
\end{displaymath}
The ${\bf Z}_2$ acts by conjugating the matrices and flipping the
sign of each $\theta$.  Thus, the ${\bf Z}_2$-invariant states
are of the form
\begin{displaymath}
\left[ \begin{array}{cc}
       0 & b_x \\ 0 & 0 \end{array} \right] \theta_1 \: + \:
\left[ \begin{array}{cc}
       0 & b_y \\ 0 & 0 \end{array} \right] \theta_2
\end{displaymath}
modulo states of the form
\begin{displaymath}
\left[ \begin{array}{cc}
       0 & d \\ 0 & 0  \end{array} \right] \theta_1 \: - \:
\left[ \begin{array}{cc}
       0 & a \\ 0 & 0  \end{array} \right] \theta_2
\end{displaymath}
so the space of physical degree one states in the orbifold
has dimension $2-2=0$.

Finally, from section~\ref{dxdy} recall the physical degree two states
on the covering space ${\bf C}^2$ have the form
\begin{displaymath}
\left[ \begin{array}{cc}
       0 & 0 \\ c & 0  \end{array} \right] \theta_1 \theta_2 
\end{displaymath}
The ${\bf Z}_2$ action conjugates the matrix as above,
and acts by a sign on each $\theta$, leaving the product
$\theta_1 \theta_2$ invariant.  This action acts nontrivially on $c$,
so that the only ${\bf Z}_2$-invariant state is defined by the
zero matrix.  Thus, the space of physical degree two states in the
orbifold has dimension zero.

Thus, we see that in order for the physical spectrum to match
Ext groups between the sheaves generated by our ansatz, we must require
\begin{displaymath}
\mbox{dim } \mbox{Ext}^n_{ [ {\bf C}^2 / {\bf Z}_2 ] }
\left( D_x, D_y \right) \: = \: 0 \: \mbox{ for all } n
\end{displaymath}
and this is a true statement, as mentioned previously in discussions of the
McKay correspondence in section~\ref{mckay}.

\section{Conclusions}

In this paper we have shown how one can encode Higgs vev
data inside sheaves, by presenting a mathematical ansatz
for encoding Higgs vevs in sheaves, and checking that ansatz
by showing that massless boundary Ramond sector spectra
in open strings between D-branes with Higgs vevs are correctly
counted by Ext groups between the sheaves generated by our ansatz.

These computations resolve several interesting puzzles.
First, our methods give us an on-shell physical interpretation
of many sheaves which did not previously have an understanding
in terms of on-shell D-branes.  This is an important step towards
making derived categories useful for physics.

Second, in the context of orbifolds, our methods have allowed us
to give a very general understanding of why {\it e.g.} Hilbert
schemes appear when describing classical Higgs moduli spaces
on D-branes in orbifolds, and how the McKay correspondence
as formulated in \cite{bkr} is consistent with physics. 
The nonreduced schemes counted by Hilbert schemes and generated
in \cite{bkr} correspond, via the correspondence in this paper,
to D-branes with nilpotent Higgs vevs, as occur in orbifolds.

Finally, our methods allow us to further clarify how
the idea that string orbifolds are strings on quotient stacks \cite{qs}
can be consistent with the fact that classical Higgs moduli spaces
of D-branes in orbifolds see resolutions of quotient spaces.
As outlined in \cite{qs}, and significantly amplified here,
exceptional divisors in classical Higgs moduli spaces,
generated by nilpotent Higgs vevs on the D-brane, are equivalent to 
$G$-equivariant nonreduced schemes, or equivalently,
nonreduced schemes on the quotient stack.  
The fact that D-branes in orbifolds ``see'' a resolution of the
quotient space corresponds to the fact that structure sheaves
of nonreduced schemes in quotient stacks can be stable at
fixed points of the group action.  Thus, it is completely consistent 
for D-branes to ``see'' a resolution of the quotient space
at the same time that the strings themselves propagate on a
quotient stack, and not a resolution of the quotient space.

\section{Acknowledgements}

E.S. would like to thank T.~Gomez for many useful discussions
of \cite{del} and for the work that led directly to this paper,
namely \cite{tomasme}.  E.S. would also like to thank S.~Hellerman
and J.~McGreevy for discussions on topics related to this paper.
R.D. is partially supported by NSF grant DMS 0104354 and by Focused
Research Grant DMS 0139799.
S.K. is partially supported by NSF grants DMS-02-96154, DMS-02-44412,
and NSA grant MDA904-03-1-0050.  E.S. is partially supported by NSF grant
DMS 02-96154.

\appendix

\section{Proof of computation of Ext groups}   \label{pfs}

In this appendix we shall outline a proof of the statement
that the BRST cohomology calculation described earlier in this paper
always gives the same result as Ext groups.
When comparing to open string spectra, we assume that
$TX|_S$ splits holomorphically into
$TS \oplus {\cal N}_{S/X}$, so that all Higgs fields in the usual
sense of algebraic geometry correspond to string modes,
and so that the spectral sequences that formed an important
part of \cite{ks,kps,cks,merev} all trivialize.
However, this technical assumption is not needed for the
statements in this appendix, only for the application of the
results in this appendix to open string spectra.

\subsection{Generalities on Higgs bundles}
\label{sec:genhiggs}
Let $Z$ be a variety and $E$ a vector bundle on $Z$.  Our main application
is to the case when $Z$ is a submanifold of a Calabi-Yau manifold $X$ and $E$ is
the normal bundle to $Z$ in $X$.  In this section, we give a definition
and standard properties of Higgs bundles.  

\begin{defn}
An {\em $E$-valued Higgs bundle\/} on $Z$ is a vector bundle $V$ on $Z$
together with a symmetric map
\[
\Phi:V\to V\otimes E.
\]
\end{defn}
In the above definition, symmetric means that the natural map
$\Phi\wedge \Phi:V\to V\otimes \Lambda^2E$ is 0.

Earlier in section~\ref{mathansatz} we briefly outlined how Higgs
fields can be used to deform sheaves to new sheaves in the total
space of the normal bundle.  Let us take a moment to 
describe that construction more systematically.

Note that $\Phi$ induces an action of $E^*$ on $V$:
\[
E^*\otimes V\to V.
\]
In turn this gives an action of the tensor algebra
$\bigotimes(E^*):=
\bigoplus_n(E^*)^{\otimes n}$ on $V$.  The symmetry of $\Phi$ is equivalent
to the condition that this action factors through the symmetric algebra
$\mathrm{Sym}(E^*)$:
\[
\mathrm{Sym}(E^*)\otimes V\to V.
\]  

Recall that $E=\underline{Spec}_Z
\mathrm{Sym}({\cal O}(E^*))$, where ${\cal O}(E^*)$ denotes the 
sheaf of sections of $E^*$.  Then the action of $\mathrm{Sym}(E^*)$ on $V$
induces the structure of an ${\cal S}(E)=\mathrm{Sym}({\cal O}(E^*))$ 
module on ${\cal O}(V)$.  The Spec construction applied to ${\cal O}(V)$
with this ${\cal S}(E)$ module structure determines a sheaf
$\cV$ of ${\cal O}_E$ modules, {\it i.e.} a sheaf
of modules on the total space of $E$.  Here ${\cal O}_E$ is the sheaf of
holomorphic functions on the total space of $E$.  
In this way, we associate an ${\cal O}_E$ module $\cV$ to the bundle $V$ on 
$Z$ together with the Higgs field $\Phi$.

Let $\cV$ now be any sheaf of ${\cal O}_E$ modules, assumed finite and flat
over $Z$.  Let $\pi:E\to Z$ be the structure map.  Then $\pi_*(\cV)$
is the sheaf of sections of a vector bundle $V$ on $Z$.  The ${\cal S}(E)$
module structure on ${\cal O}(V)$ arising from the ${\cal O}_E$
module structure on $\cV$ (the inverse of the Spec construction)
induces an action of $\mathrm{Sym}(E^*)$ on $V$, hence the structure of an 
$E$-valued Higgs bundle on $V$.  This construction is inverse to the above 
construction.

We can generalize the above discussion,
enlarging our category slightly by relaxing the requirement that $V$
be a vector bundle while keeping the finiteness condition for $\cV$ over $Z$.
We call such sheaves $V$ {\em Higgs sheaves\/}.

To summarize, we have an embedding of categories $\gamma:
\mbox{Higgs}_E \rightarrow \mbox{Sh}_E$, where $\mbox{Higgs}_E$
is the category of $E$-valued Higgs sheaves on $Z$, and 
$\mbox{Sh}_E$ is the category of ${\cal O}_{E}$ modules,
{\it i.e.} the category of sheaves on the total space of $E$.
Our convention is to denote $\gamma(V)$ by $\cV$.
Also, throughout this appendix, although we will talk about
the categories above, we will only work with sheaves in $\mbox{Sh}_E$
with the property that their support is finite over $Z$,
and we will only work with Higgs sheaves in $\mbox{Higgs}_E$
that are coherent.

There is a notion of internal Hom, $\underline{\mbox{Hom}}$, 
in each of these categories.  It is crucial to note though
that these do {\it not} correspond to each other under 
the embedding of categories $\gamma$.  
Note that $\underline{\mbox{Hom}}(V,W)$ in the
category $\mbox{Higgs}_E$ is a Higgs bundle defined by the
bundle $\underline{\mbox{Hom}}_{ {\cal O}_Z }(V,W)$
with Higgs field $-\Phi \otimes 1 + 1 \otimes \Psi$,
where $\Phi$ and $\Psi$ are the Higgs fields associated to
$V$ and $W$, respectively.
Therefore,
$\underline{\mbox{Hom}}(V,W) $ in the category $\mbox{Higgs}_E$
produces a Higgs bundle whose eigenvalues are all combinations
$\lambda - \mu$, where $\lambda$, $\mu$ are the eigenvalues of
the two Higgs fields $\Phi$, $\Psi$.
On the other hand, $\underline{\mbox{Hom}}_E( {\cal V},
{\cal W})$ keeps track only of pairs $\lambda=\mu$.

\subsection{Derived functors}
\label{df}

Fix $Z$ and $E$.
Let $(\mathrm{Higgs}_E)$ be the abelian category of $E$-valued Higgs sheaves,
and let $(\mathrm{Sh}_Z)$ denote the category of sheaves of 
${\cal O}_Z$-modules.
Consider the functor
\[
\ker:(\mathrm{Higgs}_E)\to (\mathrm{Sh}_Z),\qquad \left(
\Phi:V\to V\otimes E\right)\mapsto \ker(\Phi).
\]

We need to compute the derived functor 
\[
R\ker:D^b(\mathrm{Higgs}_E)\to D^b(Z).
\]
Consider the complex
\[
V_E^\bullet:V\to V\otimes E\to V\otimes \Lambda^2E\to \cdots
\]
given by repeated wedging with $\Phi$.  This is a complex by the symmetry
assumption.

\begin{lem}
\label{derived}
\[
R\ker\left(
\Phi:V\to V\otimes E\right) = V_E^\bullet.
\]
\end{lem}

\bigskip\noindent 
{\em Proof :\/} Let us identify $Z$ with the zero
section of $E$, and let ${\cal O}_Z$ be the structure sheaf of that
zero section.

With our identifications from Section~\ref{sec:genhiggs}, the 
$\ker$ functor gets identified
with the functor $\underline{\Hom}_{{\cal O}_E}({\cal O}_Z,\cdot)$:
\begin{equation}   \label{earlypic}
\xymatrix{
\mbox{Higgs}_E \ar[rr]^{\ker} \ar[d]^{\gamma} & & \mbox{Sh}_Z 
\ar@{=}[d] \\
\mbox{Sh}_E \ar[r]^{\underline{\Hom}( {\cal O}_Z, \cdot )} &
\mbox{Sh}_E \ar[r]^{ \pi_* } & \mbox{Sh}_Z
}
\end{equation}   
Therefore $R\ker$ is identified with the derived functor
$R\underline{\Hom}_{{\cal O}_E}({\cal O}_Z,\cdot)$ (whose cohomology
sheaves are the $\underline{{\rm Ext}}^i_{{\cal O}_E}({\cal O}_Z,\cdot)$).

Note that ${\cal O}_Z$ can be resolved by the Koszul
complex on the tautological section of $\pi^* E^*$
\begin{displaymath}
\cdots \: \longrightarrow \: \Lambda^2 \pi^* E^*  \: \longrightarrow \:
\pi^* E^* \: \longrightarrow \: {\cal O}_E
\end{displaymath}
as in \cite[Section~A.1]{ks}.
Using this Koszul resolution to compute $R\ker=
\underline{{\rm Ext}}^i_{{\cal O}_E}({\cal O}_Z,\cdot)$, we get 
the claimed result.

\bigskip\noindent
With Lemma~\ref{derived} in hand, we can relate the Higgs
spectra to Ext groups.

\bigskip
If $V$ and $W$ are two $E$-valued Higgs bundles, then $\underline{\Hom}
(V,W)$ is
naturally a Higgs bundle.  Also $\Gamma(\ker\underline{\Hom}
(V,W))$ is $\Hom(V,W)$, the space of morphisms of Higgs bundles.  Here
$\Gamma:(\mathrm{Sh}_Z)\to (\mathrm{Vect})$ is the global section
functor.  We
get
\[
\mathrm{Ext}^i_{ Higgs_E }(V,W)=R^i\Gamma(R\ker\underline{\Hom}(V,W)).
\]
(So long as $W$ is locally-free, $\underline{\mbox{Hom}}$ is exact,
and hence does not need to be derived.)
This in turn can be computed by applying Lemma~\ref{derived} 
to the Higgs bundle
$\underline{\mbox{Hom}}(V,W)$, yielding
\begin{prop}
\begin{equation}
\label{ext}
\mathrm{Ext}^i_{ Higgs_E }(V,W)=R^i\Gamma(\underline{\Hom}(V,W)_E^\bullet).
\end{equation}
\end{prop}
In the next section
we compute these Ext groups.

\subsection{Relation to open string vertex operators}
\label{vertex}

Given a Higgs bundle $V$, we want to compute $R^i\Gamma(V_E^\bullet)$.  
Then we replace $V$ by the Higgs bundle $\underline{\Hom}(V,W)$ to compute
$\mathrm{Ext}^*(V,W)$ using (\ref{ext}).

We compute $R^i\Gamma(V_E^\bullet)$ by a fine resolution: 
we have a double complex
\[
C^{p,q}=V\otimes\Lambda^p E\otimes {\cal A}^{0,q},
\]
where ${\cal A}^{0,q}$ denote the smooth $(0,q)$ forms.  The differential
in the $p$ direction is given by wedging with $\Phi$, and the
differential in the $q$ direction is $\bar\partial$.  The total
complex is $D^n=\oplus_{p+q=n}C^{p,q}$, and the total differential on
$D^n$ restricted to $C^{p,q}$ is given by $d=\Phi+(-1)^q \bar\partial$.

The inclusion $V_E^\bullet\subset C^{\bullet,0}$ induces a quasi-isomorphism
$V_E^\bullet\to D^\bullet$.  This gives
\begin{equation}
\label{hyper}
R^i\Gamma(V_E^\bullet)=H^i_d\left(\Gamma(D^\bullet)\right)
\end{equation}
since the $D^n$ are fine.

When this construction is applied to $\underline{\Hom}(V,W)$, (\ref{hyper}) is
precisely the same as the BRST cohomology, as discussed in section~\ref{brst}.

\subsection{Relation to schemes}

We have computed
$\mathrm{Ext}^*_{Higgs_E}(V,W)$ as a
hypercohomology using (\ref{ext}).  This can be computed by a
hypercohomology spectral sequence, as computed for example in
Section~\ref{vertex}.  It turns out that this is closely related to
the local to global spectral sequence which formed the mathematical basis
for the computations in \cite{ks}.

Let $\pi:E\to Z$ be the projection from the total space of $E$ to $Z$.
Let $V$ and $W$ be $E$-valued Higgs bundles, consider $\underline{\Hom}
(V,W)$ with its induced $E$-valued Higgs bundle structure as before.
Let $\underline{\Hom}(V,W)^{\bullet}_{E}$ be the Koszul complex 
associated to it as
defined in Section~\ref{df}.
Let $\mathcal{V},\ \mathcal{W}$ be the coherent sheaves on $E$ associated to
$V,\ W$ by the embedding of categories $\gamma$ discussed in 
section~\ref{sec:genhiggs}.

\begin{lem}
\[
\pi_*\left(\underline{Ext}^i_E(\mathcal{V},\mathcal{W})\right)
=\underline{H}^i(\underline{\Hom}(V,W)^{\bullet}_E).
\]
\end{lem}
Here $\underline{H}^i$ is used to denote the $i^{\scriptstyle{\rm th}}$
cohomology sheaf of a complex of sheaves.

To prove the lemma, there is a commutative diagram of functors
\begin{displaymath}
\xymatrix{
\mathrm{Higgs}_E 
\ar[r]^{\underline{\mathrm{Hom}}(V,\cdot)} \ar[d]_{\gamma} &
\mathrm{Higgs}_E 
\ar[r]^{\ker} & \mathrm{Sh}_Z \ar@{=}[d] \\
\mathrm{Sh}_E 
\ar[r]^{\underline{\mathrm{Hom}}(\mathcal{V},\cdot)} &
\mathrm{Sh}_E \ar[r]^{\pi_*} &
\mathrm{Sh}_Z
}
\end{displaymath}
Note that when $V$ is the structure sheaf of $Z$ this diagram
reduces to the earlier diagram~(\ref{earlypic}).  
The commutativity says that
\[
\ker\underline{\mathrm{Hom}}(V,W)=\pi_*\underline{\mathrm{Hom}}(\mathcal{V},\mathcal{W}).
\]
which is readily verified.
In Section~\ref{df} we have already seen an implication of this after taking
global sections, namely that
$\Gamma(\ker\underline{\Hom}
(V,W))=\Hom(V,W)$.

Passing to derived categories, we get
\[
R\ker\circ R\underline{\Hom}(V,\cdot) = 
R\pi_* \circ R\underline{\Hom}(\mathcal{V},\cdot).
\]
But $R^i\pi_*\underline{\Hom}({\cal V},{\cal W})=0$ for $i>0$, since
$\underline{\Hom}(\mathcal{V},\mathcal{W})$ is finite over $Z$. Also
$\underline{\Hom}(V,\cdot)$ is exact, hence $R\underline{\Hom}(V,W)=
\underline{\Hom}(V,W)$.  Applying this
equality to $W$ and using Lemma~\ref{derived} we get
\[
\underline{H}^q(\underline{\Hom}(V,W)^\bullet_E) =
\pi_*\underline{Ext}^q(\mathcal{V},
\mathcal{W}).
\]
Note the slight abuse of notation, objects of
$D^b(\mathrm{Higgs}_E)$, are identified with objects of
the triangulated subcategory of $D^b(E)$ associated to complexes of
sheaves finite over $Z$.

The point of all this is that the local to global spectral sequence starts
with 
\[
E_2^{p,q}=H^p(E,\underline{Ext}^q({\cal V},{\cal W})))\simeq
H^p(Z,\pi_*\underline{Ext}^q({\cal V},{\cal W})),
\]
again using the vanishing of $R^i\pi_*$ for sheaves finite over $Z$.
But now we get that this is
\[
H^p(Z,\underline{H}^q(\underline{\Hom}(V,W)^\bullet_E)).
\]
This is precisely the $E_2^{p,q}$ of one of the hypercohomology spectral
sequences
\[
E_1^{p,q}=H^p(Z,\underline{\Hom}(V,W)\otimes \Lambda^q E)
\]
that can be used to compute $\mathrm{Ext}^i_E({\cal V},{\cal W})$ 
from the Higgs bundle data.

Thus, we see that open string spectra, which were seen in section~\ref{brst}
to be given by the abutment of $E_1^{p,q}$ (in the case that $TX|_S$
splits holomorphically as $TS \oplus {\cal N}_{S/X}$),
are counted by Ext groups between the sheaves generated from Higgs data
as in section~\ref{sec:genhiggs}.

\subsection{Relation between Higgs bundles and general deformations}

We believe that open string spectra with (normal bundle valued) Higgs
fields turned on are closely related to the deformations of
a general sheaf $i_*{\cal E}$ in the case where the sheaf $i_*{\cal E}$
is unobstructed to second order.  We illustrate this somewhat imprecise
statement in a special case, namely when
\begin{itemize}
\item $TX|_S\simeq TS\oplus {\cal N}_{S/X}$
\item ${\cal E}$ is a line bundle $L$ on $S$.
\end{itemize}
The splitting of the restricted tangent bundle is not as restrictive
an assumption as it may seem at first, at least if $S$ is a curve.
For example, the normal bundle sequence splits if $H^1(S,{\cal
N}_{S/X}^{\vee} \otimes TS)=0$.  
If for instance $S={\bf P}^1$ with normal bundle
${\cal O}(a)\oplus {\cal O}(-2-a)$ and $-1\le a\le 3$ this condition
is satisfied.

We now proceed under these assumptions.  We also fix a splitting of
the restricted tangent bundle since the worldsheet analysis uses such
data.

We start with the Higgs data $\Phi\in \Hom(L,L\otimes {\cal N}_{S/X})=
H^0(S,{\cal N}_{S/X})$.  This determines a first order deformation of
$S$ inside $X$.  

The splitting of the restricted tangent bundle determines an embedding ${\cal 
N}_{S/X}\hookrightarrow TX|_S$, so we get a section in $H^0(S,TX|_S)$ from
the section of ${\cal N}_{S/X}$.  But sections of the restricted tangent bundle
are precisely the first order deformations of the embedding $S\to X$.  We
learn that our deformation of $S$ arises from a deformation of the embedding
of $S$.

More precisely, letting $D={\rm Spec}\ {\bf C}[\epsilon]/(\epsilon^2)$,
our deformation of $S$ is realized as a family (over $D$) of embeddings
\[
i_D:S\times D\to X\times D
\]
whose restriction to the closed point $0\in D$ is the original embedding of
$S$ in $X$.  As usual with nilpotents in algebraic geometry, nothing else is
happening set-theoretically besides the original embedding. The deformation 
data is encoded entirely in the scheme structure of $S\times D$ and the
morphism $i_D$.  The deformation of $i_*L$ is given by the sheaf
$(i_D)_*(L\otimes {\cal O}_D)$ on $X\times D$, which is flat
over $D$ hence gives a family of sheaves on $X$ parametrized by $D$.  For
simplicity, let's denote this family by $(i_*L)_D$.
This gives a first order deformation of the D-brane.

Since $i_D:S\times D\to X\times D$ is a local complete 
intersection, the local exts are computed by a Koszul complex exactly as
in \cite{ks}.  The result is
\[
\underline{Ext}^q_{X\times D}((i_*L)_D,(i_*L)_D)=
\Lambda^q{\cal N}_{S\times D,X\times D}.
\]
Let $\pi_D:S\times D\to D$ denote the projection.  
Then we compute the family of Ext groups via the local to global 
spectral sequence beginning with
\[
E_2^{p,q}=R^p(\pi_D)_*\left(\Lambda^q{\cal N}_{S\times D,X\times D}\right),
\]
and converging to the family of Ext groups $Ext^{p+q}_{X\times D/D}\left(
(i_*L)_D,(i_*L)_D\right)$.

As an example, the family of Exts can be computed explicitly in the case of the
obstructed ${\cO}\oplus {{\cO}} (-2)$ curve considered in section~\ref{obs}.
The result for $Ext^{1}_{X\times D/D}\left(
(i_*L)_D,(i_*L)_D\right)$ is $\cO_D$ for $n> 2$ and the sheaf $\cO_D/
\epsilon\cO_D$ for $n=2$, where $n$ is the order of the obstruction,
as in section~\ref{obs}.  

In other words the open string spectrum detects
the second order obstruction but not the higher order 
obstructions.\footnote{A similar analysis done over $D_n=
{\rm Spec}\ {\bf C}[\epsilon]/(\epsilon^n)$ would detect an obstruction at
order $n$ but not at higher order.}  A similar phenomenon can occur for the 
more general deformation of $S$ parametrized by $D$ considered above.

Now let us turn to the Higgs story.  We turn on a Higgs vev
continuously by considering the family of Higgs vevs $\epsilon \Phi$,
where $\epsilon$ is temporarily a complex number.  One might expect
that this corresponds to the spectrum of a deformation of sheaves as
above, but we will see that this is only true in an unobstructed
situation.  We look at the hypercohomology sequence which starts with
\[
E_1^{p,q}=H^p(S,\Hom(L,L)\otimes \Lambda^q {\cal N}_{S/X}).
\]
Note that the differential $d_1$ 
is identically zero for all $\epsilon$, since the induced 
Higgs bundle structure on $\Hom(L,L)$ is zero: the map 
$\Hom(L,L)\to\Hom(L,L)\otimes {\cal N}_{S/X}$ is multiplication
by $\epsilon\Phi-\epsilon\Phi=0$.

In short, much detailed information about the deformation is lost.

Let us illustrate with the ${\cal O} \oplus {\cal O}(-2)$ curve.
If we now mod out by $\epsilon^2$ and interpret the spectrum as an
$\cO_D$-module, we get for the family of $Ext^1$'s the module
$\cO_D$.  This agrees with the sheaf computation for $n>2$ but
{\em not\/} for $n=2$, the only case where the curve is obstructed at
second order.  This gives a precise meaning to our comments at
the beginning of this section.  Similar comments apply to the
more general deformation of $S$ parametrized by $D$ considered above.


\begin{thebibliography}{199}

\addcontentsline{toc}{section}{References}

\bibitem{hm} J. Harvey and G. Moore, ``On the algebras of BPS states,''
Comm. Math. Phys. {\bf 197} (1998) 489-519, {\tt hep-th/9609017}.

\bibitem{ks} S. Katz and E. Sharpe, ``D-branes, open string vertex
operators, and Ext groups,'' Adv. Theor. Math. Phys. {\bf 6} (2003) 979-1030,
{\tt hep-th/0208104}.


\bibitem{kps} S. Katz, T. Pantev, and E. Sharpe, ``D-branes, orbifolds,
and Ext groups,'' {\tt hep-th/0212218}.

\bibitem{cks} A. Caldararu, S. Katz, and E. Sharpe, ``D-branes, B fields,
and Ext groups,'' {\tt hep-th/0302099}.

\bibitem{merev} E. Sharpe, ``Lectures on D-branes and sheaves,''
{\tt hep-th/0307245}.


\bibitem{paulron} P. Aspinwall and R. Donagi,
``The heterotic string, the tangent bundle, and derived categories,''
Adv. Theor. Math. Phys. {\bf 2} (1998) 1041-1074,
{\tt hep-th/9806094}.

\bibitem{medc} E. Sharpe, ``D-branes, derived categories, and
Grothendieck groups,'' Nucl. Phys. {\bf B561} (1999) 433-450,
{\tt hep-th/9902116}.

\bibitem{mikedc} M. Douglas, ``D-branes, categories, and
${\cal N}=1$ supersymmetry,'' J. Math. Phys. {\bf 42} (2001) 2818-2843,
{\tt hep-th/0011017}.

\bibitem{paulalb} P. Aspinwall and A. Lawrence, ``Derived categories
and zero-brane stability,'' JHEP 0108 (2001) 004, {\tt hep-th/0104147}.






\bibitem{tomasme} T. Gomez and E. Sharpe, ``D-branes and scheme theory,''
{\tt hep-th/0008150}.

\bibitem{dgm} M. Douglas, B. Greene, and D. Morrison, ``Orbifold resolution
by D-branes,'' Nucl. Phys. {\bf B506} (1997) 84-106,
{\tt hep-th/9704151}.

\bibitem{bkr} T. Bridgeland, A. King, M. Reid, ``Mukai implies McKay,''
J. Amer. Math. Soc. {\bf 14} (2001) 535-554, {\tt math.AG/9908027}.

\bibitem{qs} E. Sharpe, ``String orbifolds and quotient stacks,''
Nucl. Phys. {\bf B627} (2002) 445-505, {\tt hep-th/0102211}.


\bibitem{calin} C. Lazaroiu, ``Generalized complexes and string
field theory,'' {\tt hep-th/0102122}.

\bibitem{del} R. Donagi, L. Ein, and R. Lazarsfeld, ``A non-linear deformation
of the Hitchin dynamical system,'' {\tt alg-geom/9504017},
a.k.a. ``Nilpotent cones and sheaves on K3 surfaces,'' pp. 51-61 in
{\it Birational Algebraic Geometry}, Contemp. Math. 207,
Amer. Math. Soc., Providence, Rhode Island, 1997.

\bibitem{tonysb} T. Pantev, lectures at KITP miniprogram on
``Geometry, Topology, and Strings,'' summer 2003,
available at
{\tt http://online.kitp.ucsb.edu/online/mp03/pantev1}.



\bibitem{fulton} W. Fulton, {\it Intersection Theory}, second edition,
Springer, Berlin, 1998.


\bibitem{mohri} K. Mohri, ``K\"ahler moduli space for a D-brane at
orbifold singularities,'' {\tt hep-th/9806052}.

\bibitem{itonakajima} Y. Ito and H. Nakajima, ``McKay correspondence
and Hilbert schemes in dimension three,'' {\tt math.AG/9803120}.


\end{thebibliography}
\end{document}